\documentclass[acmsmall]{acmart}

\usepackage{graphicx}
\usepackage{amssymb}
\usepackage{verbatim}
\usepackage{url}
\usepackage{tikz}
\usepackage{array}
\usepackage{subfig}
\usepackage{pgfplots}
\pgfplotsset{compat=1.14}
\usepackage{graphicx}
\usepackage{enumitem}
\usepackage{lscape}
\usetikzlibrary{arrows,shapes,positioning,shadows,trees}

\usepackage{forest}
\usetikzlibrary{arrows.meta}

\tikzstyle{arrow} = [thick,->,>=stealth]

\newtheorem{trade-off}{Trade-off}

\newtheorem{definition}{Definition}
\newtheorem{proposition}{Proposition}

\AtBeginDocument{%
  \providecommand\BibTeX{{%
    \normalfont B\kern-0.5em{\scshape i\kern-0.25em b}\kern-0.8em\TeX}}}

\setcopyright{acmcopyright}
\copyrightyear{2018}
\acmYear{2018}
\acmDOI{10.1145/1122445.1122456}

\acmJournal{TOS}
\acmVolume{00}
\acmNumber{00}
\acmArticle{000}
\acmMonth{00}

\begin{document}

\title{A Survey on Large Scale Metadata Server for Big Data Storage}

\author{Ripon Patgiri}
\email{ripon@cse.nits.ac.in}
\orcid{0000-0002-9899-9152}
\author{Sabuzima Nayak}
\email{sabuzimanayak@gmail.com}
\affiliation{%
  \institution{National Institute of Technology Silchar}
  \streetaddress{Department of Computer Science \& Engineering}
  \city{Silchar}
  \state{Assam}
  \country{India}
  \postcode{788010}
}








\renewcommand{\shortauthors}{Patgiri and Nayak}

\begin{abstract}
Big Data is defined as high volume of variety of data with an exponential data growth rate. Data are amalgamated to generate revenue, which results a large data silo. Data are the oils of modern IT industries. Therefore, the data are growing at an exponential pace. The access mechanism of these data silos are defined by metadata. The metadata are decoupled from data server for various beneficial reasons. For instance, ease of maintenance. The metadata are stored in metadata server (MDS). Therefore, the study on the MDS is mandatory in designing of a large scale storage system. The MDS requires many parameters to augment with its architecture. The architecture of MDS depends on the demand of the storage system’s requirements. Thus, MDS is categorized in various ways depending on the underlying architecture and design methodology. The article surveys on the various kinds of MDS architecture, designs, and methodologies. This article emphasizes on clustered MDS (cMDS) and the reports are prepared based on a) Bloom filter-based MDS, b) Client-funded MDS, c) Geo-aware MDS, d) Cache-aware MDS, e) Load-aware MDS, f) Hash-based MDS, and g) Tree-based MDS. Additionally, the article presents the issues and challenges of MDS for mammoth sized data.
\end{abstract}

\ccsdesc[500]{Computer systems organization~Embedded systems}
\ccsdesc[300]{Computer systems organization~Redundancy}
\ccsdesc{Computer systems organization~Robotics}
\ccsdesc[100]{Networks~Network reliability}

\keywords{Metadata, Metadata Server, Metadata Management, Datacenter, Database, Clustered Metadata Server, Distributed Metadata Server, File Systems, Big Data}

\maketitle

\section{Introduction}
Big Data is the big revenue of modern IT industries. Currently, we are in the age of Big Data. Emerging technology such as Cloud Computing \cite{BUYYA, ZISSIS}, and IoT \cite{GUBBI,KHAN1} have paved the path for the emergence of Big Data. Every devices are made smart by embedding sensors inside the device. These devices are connected to every other device through the Internet. Constantly, these devices are generating data encouraged by Cloud technology \cite{BigData,Fernando}. These data are stored in the Cloud. The user of the smart devices are also availing quick and good quality of service due to another new technology called Edge Computing \cite{KHAN,AHMED} and Edge Analytics. It is making the Cloud computing utilities and services closer to users. Moreover, it performs fast processing and fast application response. Billions of people using multiple smart devices, desire to remain connected with close people round the clock, and constantly producing data which produces a huge data silo, known as Big Data \cite{RP16}. 

Metadata is data about data. Metadata delivers information about the data. Now the attention has migrated from data to metadata, i.e., from data server to metadata server. Big Data storage is a distributed system consisting of hundred of thousands of systems \cite{MDS,dMDS}. When a user request for a data, all the systems are searched. The size of data is huge, hence, finding the data is very time consuming. However, the size of a metadata is independent of its data size. Accumulated metadata size becomes huge when the total data size grows. Big Data storage technology focuses more on efficient retrieval of metadata because data are stored in different locations. Hence, Big Data technology is currently focusing more on designing of efficient MDS.

MDS is a server that serves the metadata request of the client. The MDS is decoupled from the data server \cite{Impact}. To access data, first, a client sends a request to MDS. The metadata server quickly searches for the metadata and sends the locations of the data in the Big data storage system. The client sends another request to the data server specifically to the system having the required data. This process saves the searching time of the data server not having the data. This paper focuses on MDS and its role in the Big Data storage.

\subsection{Motivation}
There are some research questions (RQ) to be addressed which are listed below-
\begin{description}
\item[RQ1:] Why does the metadata cohere with the data?
\item[RQ2:] Why is the MDS decoupled from data server?
\item[RQ3:] Why does scalability of the MDS so important?
\item[RQ4:] How does the MDS impact on performances of a Big Data storage system?
\item[RQ5:] What are the significant method to devise new MDS?
\item[RQ6:] What are the state-of-the-art MDS technology?
\item[RQ7:] What are the issues and challenges of MDS?
\end{description}

The research questions (RQ) motivate the article to explore in-depth insight into the MDS. The \textbf{RQ1} and \textbf{RQ2} gives the overall view of metadata, MDS and data server. The \textbf{RQ3} defines the essences of the large scale MDS to perform millions of metadata operations per unit time. The \textbf{RQ4} exploits the impact of MDS in a storage system. The \textbf{RQ5} explores the methods used to enhance ultra-large file system. The \textbf{RQ6} provides the most modern MDS design techniques. And finally, the \textbf{RQ7} gives the issues and challenges of the MDS.

\subsection{Organization}
The article is structured as follows: Section \ref{BDS} sketches on Big Data storage system. Section ~\ref{md} highlights the various metadata operations. Besides, the section discusses on various types of MDS. Section \ref{MS} discusses about MDS. Also, the section classifies the types of MDS. Section \ref{method} explores the methods used in designing MDS. Section ~\ref{mmds} surveys large scale MDS for Big Data storages. Section ~\ref{issue} discloses the various issues in designing MDS. Section \ref{dis} discusses various aspects of designing issues of MDS. Finally, the article is concluded with section \ref{con}.


\section{Big Data Storage System}
\label{BDS}
Big data is voluminous data of size petabyte, Exabytes or beyond. Big Data requires multiple systems to store such huge volume of data. Distributed file system consists of multiple systems that agreed to work as a single entity to provide services to the users. They share data, computation and resources for providing quality of services. Some examples \cite{Ma} of earlier distributed file systems are network file system (NFS) \cite{NFS}, Google file system (GFS) \cite{GFS03}, Lustre \cite{lustre}, and Hadoop distributed file system (HDFS) \cite{HDFS}. HDFS is implemented to store medical images in cloud \cite{YANG}. In this section, among the earlier distributed file systems only HDFS is explained to give a brief understanding about distributed file systems.

\subsection{Hadoop Distributed File System}
HDFS design is modeled on the GFS. HDFS is the fundamental distributed storage used by the Hadoop applications. Hadoop \cite{hadoop} is a large-scale distributed system having a batch processing infrastructure. Hadoop has large scalability. It follows a master-slave architecture. The master node is called NameNode and the slave nodes are called DataNodes. The client interacts with the NameNode and NameNode takes the responsibility of managing the DataNodes. The Clients stores data in HDFS and data storage and management is handled by the HDFS. However, this master-slave architecture has the single point of failure issue. HDFS uses low commodity hardware, hence, the probability of failure of the master node i.e. NameNode is high. But this architecture separates the metadata and data which increases efficiency. The data are partitioned into blocks and stored across multiple DataNodes. This provides the storage scalability. Big data can be stored by storing different blocks in different DataNodes. This feature has enabled the HDFS to be scalable and provide storage to huge volume of data. The NameNode stores the metadata of the files. The Client sends request to NameNode for data. The NameNode gives the location of the DataNode. After receiving the locations the Client directly contacts the DataNode for the data. Instead of searching all DataNodes for the presence of data. The NameNode only searches for the metadata of the data. 

\subsection{Classification of Storage System Architecture}
The architecture of storage system is classified into three types, namely, Direct-Attached Storage (DAS), Network Attached Storage (NAS), and Storage Area Network (SAN) \cite{taxonomy}.

\begin{enumerate}
\item \textbf{Direct-Attached Storage}
DAS \cite{DAS} has very simple architecture  where every server has its own storage. DAS has an easy management and more appropriate for local services. It uses IDE/SCSI for transmission. It stores the data in track and sectors. For fault tolerance it uses RAID. However, it is not scalable and only capable of providing services to local users. 

\item \textbf{Network Attached Storage}
NAS \cite{NAS} has the storage separated. Storage is connected using an Ethernet switch. It helps to increase the scalability. Storage is accessed using TCP/IP protocol. NAS has an easy disaster recovery system. The data is stored using the shared files technique. NAS is fault tolerance by replicating the data. NAS is capable to provide services to long distance users. However, performance of NAS is not high and it has less efficient than SAN.  

\item \textbf{Storage Area Network}
In SAN the storage is separated and connected using fiber switch. Fiber switch reduces the data access time. Hence, increases the performance of SAN which is more compared to both NAS and DAS. However, it has scalability issues because fiber channels are expensive and not possible for long distance connection. SAN storage uses blocks to store data. For fault tolerance it uses RAID.
\end{enumerate}

\subsection{Types of Big Data Storage}

\begin{enumerate}
    \item \textbf{Object storage:} Big data use object storage in case the data are stored using objects \cite{taxonomy,object}. In object storage the data objects are stored in the devices as a set of logically connected bytes. It has methods for accessing the data. The object consists of both metadata and data. Object storage is most appropriate for storage of unstructured data of Big Data. Some examples of implementation of object storage is Amazon S3 \cite{amazons3} and Openstack Swift \cite{osswift}. 
    
    \item \textbf{Block Storage:} In case of block storage system \cite{taxonomy,block}, the data are stored in blocks of storage media. This storage type depends on the network of storage area. This reduces its scalability. This storage only stores raw data. Some examples of implementation of block storage are Amazon Elastic block store (EBS) \cite{amazonebs} and Openstack Cinder \cite{openstack}.
    
    \item \textbf{Cloud Storage:} Another type of storage is cloud storage \cite{taxonomy,zeng}. Cloud storage stores the data on multiple servers. It uses file system, object storage, and hybrid storage system. The Cloud storage architecture consists of three layers, namely, user application,  storage management and storage resources. The user application layer act as an interface between the  virtual storage media and user. The storage management layer is responsible for the virtualization of the storage space. The layer manages the storage. Virtualization helps to appear the storage as a single unit, whereas, actually the data are stored in multiple servers. The storage resources layer has the responsibility to store the data in storage devices. It uses file system, object storage, and hybrid storage system. The advantages of cloud storage are  availability, scalability, fault tolerance, security, and so on \cite{Huo, Wu}. 
\end{enumerate}

\subsection{Database}
Big Data is not completely structured, it also consists of semi-structured and unstructured data. Hence, RDBMS is incapable to handle Big Data. NoSQL i.e. not only SQL is defined as non-relational, distributed, open-source and horizontal scalable database. NoSQL \cite{taxonomy,nosql} is implemented in parallel and distributed computing. Therefore, NoSQL database is more appropriate for Big Data. The advantages of NoSQL are scalability, high availability, reliability and fault-tolerance. The NoSQL is classified mainly into four types, namely, key-value database, column-oriented database, document-oriented database and graph database \cite{taxonomy, nosql}.

\subsubsection{Key-Value Database}
In key-value NoSQL database, the data are represented as a key and value. The key is unique to identify the data and the value is the data. Hashing is used to produce unique key. Some database does not restrict the data type in the value part of the format. Hence, key-value database also supports unstructured Big Data. Some examples that implements key-value database is Cassandra \cite{cassandra}, Berkeley DB \cite{berkeley} and Redis \cite{Redis}. The advantages are high scalability, availability, performance, simple design and efficient data management. 

\subsubsection{Column-Oriented Database}
In column-oriented NoSQL database, each attribute (column) of the data are stored separately. It means all the data having the same attribute are stored together. It speedup the processing of both structured and unstructured data. Some examples are HBase \cite{HBase}, and BigTable \cite{bigtable}. The advantages are high scalability, high performance and  easy design. However, it has moderate flexibility.

\subsubsection{Document-Oriented Database}
Document-oriented database stores all documents of same type in the same dataset. It is a schema free and semi-structured database. Some examples are MongoDB \cite{mongodb}, CouchDB \cite{couchdb} and HyperDex \cite{hyperdex}. The advantages of document-oriented databases are scalable and fault-tolerance to large-scale computation. 

\subsubsection{Graph Database}
A graph database stores the data representing graphs. The graph is semi-structured data which is connected and complex. Graph database is schema free and uses directed adjacency lists to represent the data. Some examples are Pregel \cite{pregel}, Giraph \cite{giraph}, PowerGraph \cite{powergraph}. The graph database uses traversal for query operations. It can easily scale horizontally. It also provide graph partition functionality. Big data consists of structured, semi-structured and unstructured data. Hence, the database should be capable of storing all types of data. In addition, the database has to provide efficient processing and query services while accessing huge volume of data. 

\subsection{Query Systems}
A query system is also important for Big data storage system. Because a Big Data storage system receives requires at a frequency of millions per second. Moreover, the query system has to scan petabyte or Exabyte of data. Hence, an efficient query format and fast processing query system are essential. Dremel \cite{Dremel} is an interactive ad-hoc query engine. It combined columnar data layout and multi-level execution trees to execute millions frequence Big Data requests. It pushes the request to a serving tree similar to web search and rewrite at every step. To obtain the response for the query the replies generated at the lower level of the tree is combined. Dremel has a query dispatcher. Its responsibility is to schedule query based on priorities, load balancing and also providing fault tolerance. Dremel is also a multi-user system where multiple queries are processed parallely. For storage representation, Dremel uses column-strip. It helps to read less data from hard disk. It does cheaper compression to reduce CPU cost. However, Dremel spends more time on decompression. HDFS uses Hive \cite{hive} for the query to its file system. Hive is an open-source data warehousing solution. It has HiveQL which is a declarative language similar to SQL. HiveQL translates the query to map-reduce task. And, then these tasks are executed in Map-Reduce. However, this increases the delay in response even for small queries. Hive has flexibility for new schema design. It is possible because Hive stores the schema separately from data. HBase also implements Hive. Similar to the Hive is another query engine called Impala \cite{impala}. It has low latency in contrast to Hive. Impala is flexible and uses standard components (eg. HBase, HDFS) of Hadoop. Moreover, it is capable of reading widely-used file formats. To reduce latency it deploys daemon on Hadoop systems. Daemons are responsible for receiving client request, selecting system for query processing and load balancing.

\section{Metadata}
\label{md}
Generally, the most widely known definition of metadata is defined in Definition \ref{def0}.
\begin{definition}\label{def0}
``Metadata is data about data''.\end{definition} Nevertheless, there are many definitions of metadata. NISO defines metadata in Definition \ref{def1}.
\begin{definition}\label{def1}
NISO defines metadata in a more formal way-  ``Metadata is structured information that describes, explains, locates, or otherwise makes it easier to retrieve, use, or manage an information resource\cite{UM17}.''\end{definition} 
Metadata provides information about data. The data does not have its own existence without metadata. Another inetersiting definition of metadata is defined in Definition \ref{def2}. \begin{definition}\label{def2}
The metadata tells ``what to access, where to access, and how to access''.\end{definition} 
Metadata is integral part of file system. NISO again defined metadata in Definition \ref{def3}.
\begin{definition}
\label{def3}
``Metadata is key to ensuring that resources will survive and continue to be accessible into the future'' ~\cite{UM17}\end{definition} 
NISO again describes the daily uses of metadata in our life as-
\begin{quote}
``... locate video on YouTube, manage finances through Quicken, connect with others via email, text, and social media and store lengthy contact lists on their mobile devices. All of this content comes with metadata - information about the item’s creation, name, topic, features, and the like. Metadata is key to the functionality of the systems holding the content, enabling users to find items of interest, record essential information about them, and share that information with others \cite{UM17}.'' 
\end{quote}

Some examples of metadata are data size, data name, time and date of creation, date of last updated, author of data etc. The metadata is inevitable parts of data storage. The purpose of metadata is to define the access mechanism of data on storage media. Without metadata, a data cannot be accessed or the data become orphan. Therefore, metadata strongly coheres with data. The metadata enhances the read/write performance of data storage. There are many metadata operations, however, some MDSs do not comply with POSIX standards depending on the underlying architecture.

\subsection{Metadata Schema}
Attribute is defined as the various parameters of a metadata. For example, filename, file size, or date of creation. Metadata schema or schema \cite{UM04} is a set of metadata attributes (elements) formulated for specific purposes. Mostly schema gives a format of describing the data. The meaning of attributes is called semantics of the schema. The schema also contains rules for the value of the attribute. For example, file size is a number with byte in abbreviation (eg. MB, KB). Metadata schema also contains syntax rules for attributes and its value. Examples of some metadata schema are XML (Extensible Mark-up Language) \cite{xml}, Standard Generalized Mark-up Language (SGML) \cite{sgml}, Dublin Core \cite{dublin}, and  Metadata Object Description Schema (MODS).

\subsection{Types of Metadata}
Metadata is classified into three types, namely, descriptive, structural and administrative \cite{UM17, UM04}.
\begin{enumerate}
    \item Descriptive metadata: This metadata describes a data. The attributes help in understanding the data. Some examples of metadata attributes are title, author, keywords etc. This metadata is used commonly for display, discovery and interoperability of the data.
    \item Structural metadata: It gives the information regarding the grouping of the data. For example, when displaying the list of files in the system, the files are sorted by last updated file, alphabetic ordering etc. It is commonly used for navigation between data. 
    \item Administrative metadata: It records information for easy management of data. It is further classified into technical metadata, preservation metadata and rights metadata. 
    \begin{enumerate}
        \item Technical metadata: It  helps in decoding the files. Some examples of attributes are file size, file type and data creation date. Technical metadata is commonly used for digital object management, interoperability,  and preservation of data.
         \item Preservation metadata: It helps in the management of data for long durations.  Examples of attributes are preservation event and checksum. It is also commonly used for digital object management, interoperability,  and preservation of data.
        \item Rights metadata: It saves the intellectual property rights information applicable on the data. Examples of attributes are rights holder, license terms and copyright status. Commonly used for digital object management and interoperability,

    \end{enumerate}
\end{enumerate}

\subsection{Metadata operations}
\begin{figure}[ht]
\centering
\scalebox{0.6}{
\newcolumntype{C}[1]{>{\centering}p{#1}}
\begin{forest}
 for tree={
  		if level=0{align=center}{
    		align={@{}C{30mm}@{}},
  		},
  		grow=east,
  		draw,
  		font=\sffamily\bfseries,
  		edge path={
    		\noexpand\path [draw, \forestoption{edge}] (!u.parent anchor) -- +(3mm,0) |- (.child anchor)\forestoption{edge label};
  		},
  		parent anchor=east,
  		child anchor=west,
  		l sep=10mm,
	}
	 [POSIX-compliant\\ Metadata \\ Operations
	 [misc
      [statfs]
      [truncate]
      [getlayout]
     ]
    [link
     	[unlink]
     	[symlink]
        [readlink]
     ]
     [prime
        [rename]
        [lookup]
        [creat]
    ]
     [attr
         [setattr]
         [getattr]
     ]
     [dir
        [rmdir]
        [readdir]
        [mkdir]
     ]
  ]
\end{forest}
}
\caption{\textbf{POSIX-compliant metadata operations. $misc$ is miscellaneous \protect\cite{Cha,posix}}}
\label{metataxo}
\end{figure}
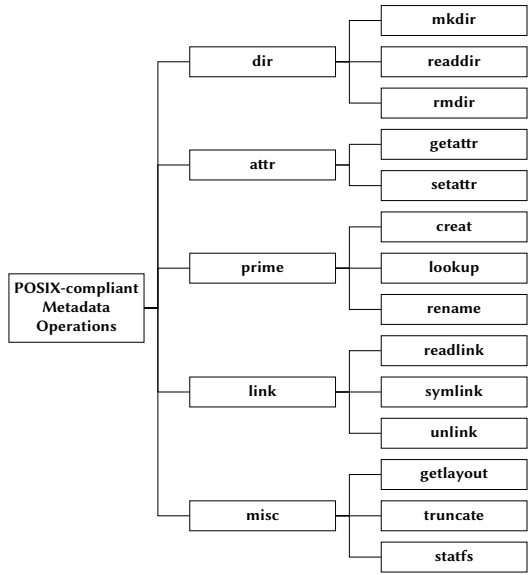

\begin{figure}
\centering
\begin{forest}
  for tree={
    align=center,
    parent anchor=south,
    child anchor=north,
    font=\sffamily,
    edge={thick, -{Stealth[]}},
    l sep+=10pt,
    edge path={
      \noexpand\path [draw, \forestoption{edge}] (!u.parent anchor) -- +(0,-10pt) -| (.child anchor)\forestoption{edge label};
    },
    if level=0{
      inner xsep=0pt,
      tikz={\draw [thick] (.south east) -- (.south west);}
    }{}
  }
  [Non-POSIX Metadata Operations
    [create]
    [open]
    [readdir]
    [lookup]
    [move]
    [rename]
    [delete]
  ]
\end{forest}
\caption{\textbf{Standard metadata operations}}
\label{metataxo1}
\end{figure}
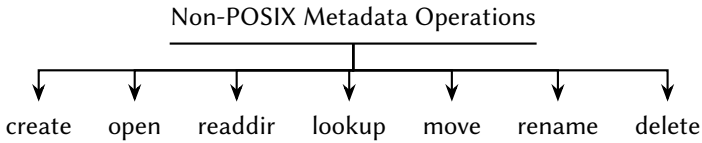

The metadata operations depend on MDS architecture. Also, the number of metadata operations varies depending on the underlying MDS architecture and design. The prime focus of MDS design is to make an in-memory MDS using a suitable data structure; HashMap, for instance. Similarly, B$^+$Tree. Notwithstanding, some modern file system does not comply with POSIX-standard metadata operations. For example, HDFS \cite{HDFS15}. On the contrary, CephFS \cite{SW04} and CFFS \cite{CFFS} comply with the POSIX - standard. The POSIX-compliant metadata operations are broadly categorized into five, namely, dir, attr, prime, link and miscellaneous. The requirements of metadata operations are more in POSIX-compliant. The POSIX-compliant MDS provide all operations. On the other hand, the non-POSIX standard metadata operations provides a few metadata operations. Important metadata operations are explained in this section. 
\begin{itemize}
    \item POXIS-compliant metadata operations:
The five classified types of POXIS-compliant metadata operations are $dir$, $attr$, $prime$, $link$, and $misc$ (miscellaneous). Figure \ref{metataxo} listed the common POXIS-compliant metadata operations.
$dir$ operation consists of the operations that performs on the directory of the file system. $mkdir$ operation creates a new directory. It takes a path and mode (directory access permission) as arguments. $readdir$ operation reads a directory. It takes the path of the directory as argument. $rmdir$ deletes a directory. It also takes the path of the directory as argument. $attr$ perform operations on attributes of the directory. Attributes refer to the different parameters of a file or directory. For example, the size of file, mode of a file, etc. $getattr$ returns the value of an attribute. It takes a path and attribute name as arguments. $setattr$ assigns value to an attribute. In some attribute this operation is not permitted. For example, changing the file size. $link$ operation creates links to a file. This link is a symbolic link. Symbolic link is one file pointing to another file or directory. $readlink$ reads the content of a file whose path is given as argument. $symlink$ creates symbolic links to another file or directory. It takes two paths as argument. One is the current file and the other is the target file. $unlink$ operation deletes a link from a file. It takes the target path as argument. $prime$ operations are performed on the files. $creat$ operation creates a new file or rewrite an existing file. It takes the filename and mode of the file as argument. $rename$ operation changes the file name. It takes the filename and the new name as the argument. Some important miscellaneous operations are $getlayout$, $truncate$ and $statfs$. $truncate$ operation cuts the file to the specified length. It takes the filename and the required length of the file as arguments. $statfs$ operation returns the information about a mounted file system. It takes the file path and a buffer to store the value as arguments. 

\item Non-POSIX Metadata Operations:
Some of the important Non-POSIX Metadata operations are $create$, $open$, $readdir$, $lookup$, $move$, $rename$, $delete$. Figure \ref{metataxo1} listed the common non-POXIS metadata operations. $create$ operation creates a new file or adds new content to an existing file. The operation takes a filename and file access permission as arguments. $open$ operation opens a file for editing or reading. It takes the filename and operation to be performed (e.g. read, write) as arguments. $readdir$ and $lookup$ operations are similar to $readdir$ and $lookup$ of POXIS-compliant metadata operations. $move$ operation changes the location or path of the file. It takes a filename, current path and new path as arguments. $rename$  operations are similar to $rename$ of POXIS-compliant metadata operation. $delete$ operation deletes the file or directory. It takes a filename/directory name and path as arguments.  
\end{itemize}

The performance of metadata operation may also vary due to distributed in nature, for example, querying metadata request may take $O(1)$ time complexity, but query to one MDS produces a query to another MDS in clustered MDS (cMDS); as a consequence, it causes network access and time delay \cite{dMDS}. Patgiri et al. \cite{dMDS} also emphasizes on cMDS and sMDS operations, and shows that the time complexity becomes inefficient to measure due to network accesses. The performance of a metadata operation depends on the underlying architecture of MDS. The operations of standalone MDS never encounters network access, but cMDS operation encounters frequently. In a large-scale space, the cMDS plays a vital role in serving large-sized metadata. On the contrary, the standalone MDS is ideal in case of small scale metadata.

\section{Metadata Servers}
\label{MS}
MDS is dedicated server(s) to serve metadata. The purpose of MDS is to serve information about data to clients. The MDS stores the metadata in primary memory to serve the information about the data as fast as possible. Nonetheless, the primary memory (RAM) is volatile in nature, that's why the MDS ensures the durability of the metadata. Sudden failure can cause metadata lost, therefore, metadata is replicated in several machines. However, consistency and durability of a metadata depend on MDS architecture. Metadata is the most crucial part of a file system for storage of Big Data to enhance the performance, and thus, the MDS defines scalability and performance enhancement of a file system storing Big Data \cite{dMDS,MDS}.

\subsection{Categories of Metadata Servers}
There are two categories of MDS, namely, standalone MDS and clustered MDS (cMDS). The standalone MDS is a single MDS to serve metadata. The clustered MDS is a set of MDS to serve large sized metadata. Furthermore, the cMDS is classified into two subcategories, namely, distributed MDS (dMDS) and parallel MDS (pMDS). Table \ref{differncetab} lists the difference between standalone MDS and cMDS.

\begin{table}[!ht]
\caption{\textbf{Comparison between Standalone MDS and Clustered MDS. \\ NA: Not applicable.}}
    \centering
    \begin{tabular}{|p{6cm}|p{2cm}|p{2cm}|}
    \hline
    \centering \vspace{0.1cm}\textbf{Parameter} &  \textbf{Standalone MDS} & \textbf{Clustered MDS}\\ \hline \hline
    No of systems & One & Many \\ \hline
    Bottleneck Issue & Yes & No \\ \hline
    Scalable & No & Yes \\ \hline
    Flexible (Change in schema) & Yes & Yes \\ \hline
    Single point of failure & Yes & No \\ \hline
    Latency Issue & NA & Yes \\ \hline
    Hotspot & NA & Yes \\ \hline
    Small File Problem & Yes & Maybe \\ \hline
    Metadata Management & Easy & Difficult \\ \hline
    Rate of parallelism & Low & High \\ \hline
    Synchronisation Issue & NA & Yes \\ \hline
    Load Balancing Issue & NA & Yes \\ \hline
    Appropriate for Big Data System & No & Yes \\ \hline

    \end{tabular}
    \label{differncetab}
\end{table}

\subsubsection{Standalone MDS}

\begin{figure}
     \centering
     \subfloat[][Architecture of standalone MDS]{\includegraphics[width=0.45\textwidth]{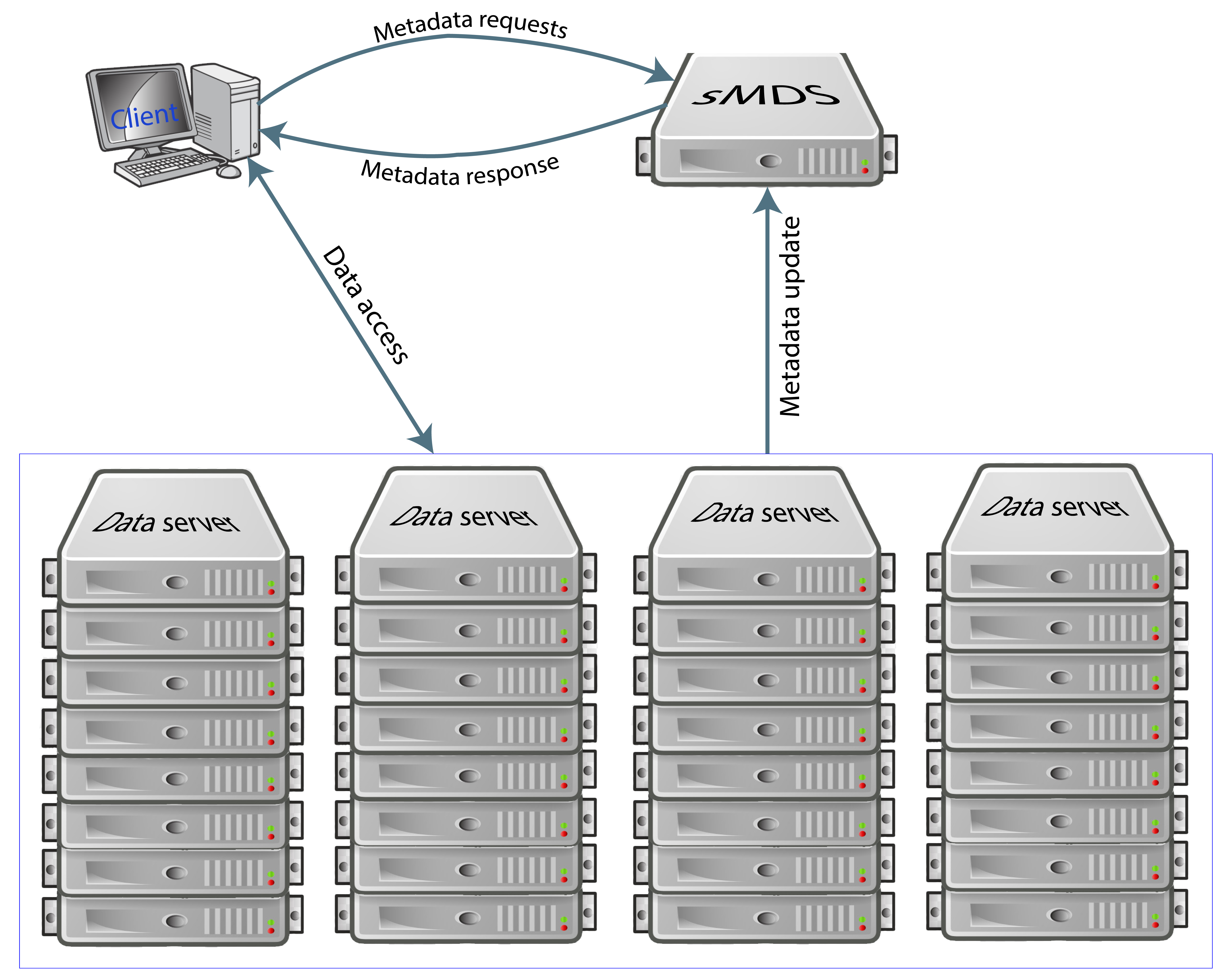}\label{MDS}} \hspace{1cm}
     \subfloat[][Architecture of clustered  MDS]{\includegraphics[width=0.45\textwidth]{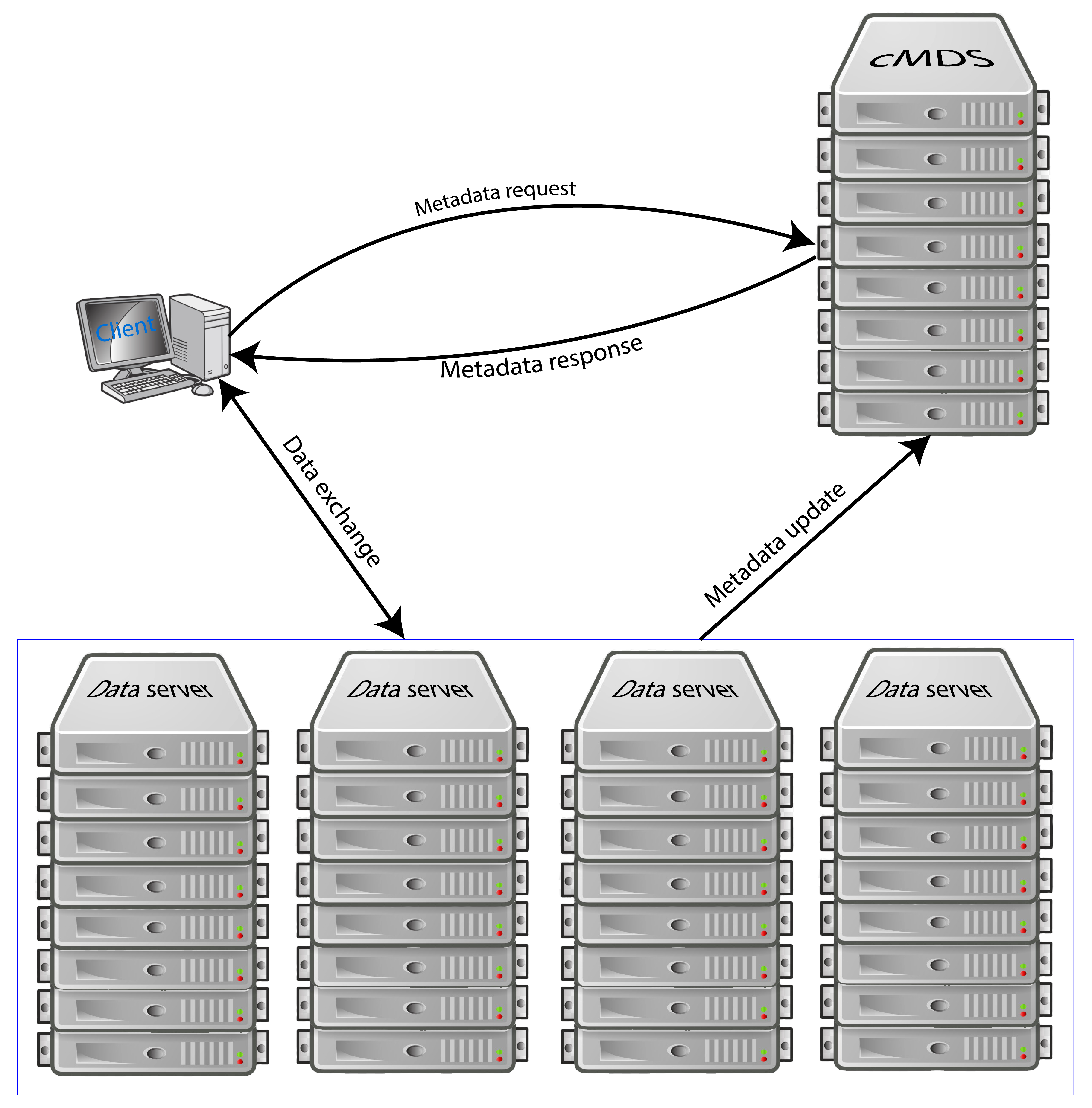}\label{cMDS}}
     \caption{Metadata Server Architecture}
     \label{steady_state}
\end{figure}

Figure \ref{MDS} illustrates the architecture of standalone MDS. 
Patgiri et al. \cite{dMDS} reported that the standalone MDS is ideal if the metadata is very less or all all metadata fits in main memory. Apparently, there is no network latency issue in standalone MDS. The standalone MDS concerns mainly on RAM, and HDD/SSD. RAM is used for suitable data structure and HDD/SSD is to create a backup and swap-out/in of metadata. The metadata is stored in HDD/SSD for permanent storage. Moreover, if metadata exceeds the RAM size, then the metadata is written to the HDD/SSD. In addition, this process is faster than accessing metadata from a remote machine through the network, but it lacks parallelism of metadata services and it experiences bottleneck. But, the machine capabilities are increasing, namely, processing power, RAM size, cache size, network speed and much more along with metadata. Therefore, the standalone MDS has been deployed in many file systems including HDFS \cite{HDFS15}, Quantcast File System (QFS) \cite{QFS13}, GFS \cite{GFS03}, and many more. 

The standalone MDS can have a limited sized of RAM. Therefore, the standalone MDS is not scalable. Patgiri \cite{MDS} calculate the metadata scalability of $r~GB$ RAM at most \[\frac{2^{10}}{m}\times n\times r \times 2^{10}~GB\] where the block size is $n~MB$, and metadata size is $m~KB$ on an average. The free memory space in a system of $4~GB$ RAM is very low because the operating system occupies nearly $1~GB$ RAM and other applications also occupy some memory. Therefore, the available RAM space becomes low. On the other hand, the RAM size comes with very large to satisfy the requirement of metadata spaces. Most of the modern servers come with $64GB$, $128GB$, $256GB$, $512GB$ and $1024GB$ of RAM spaces. Patgiri \cite{MDS} indicates that $64TB$ data can be served by $16GB$ available free RAM. It seems excellent scalability, but these $64TB$ data contributes a very large amount of users as referred to ``Vendee'' in Big Data Paradigm \cite{RP16}. These large numbers of users create millions of metadata operation requests and it is enough to slow down the system. Therefore, the scalability of MDSs becomes a prominent aspect of Big Data storage systems.

The scaling standalone MDS is prominent challenge, but sMDS cannot serve ultra large file system. For example, HDFS NameNode \cite{HDFS15}. Therefore, it is not appropriate for Big Data systems. In fact, sometimes, a few sizes of data consumes entire memory. If few GB of data can eat up the intact storage of an MDS as addressed in \cite{HAR15,HAR+15,TyrFS}, thus, the system cannot scale more data size. This situation occurs  frequently and it is the most likely probable event in standalone MDS. For instance, the worst situation happens when the data size and the metadata size of those data are approximately same. The total size of small sized files are very less to store in the storage media, but nearly equivalent to the metadata size. Hence, a set of small sized files can consume entire RAM spaces in a standalone MDS.

On the contrary, the standalone MDS pays the synchronization cost, latency issue, and re-adjustment cost for load balancing. But, it has many issues, namely, hot-standby, data recovery, bottleneck, single point of failure (SPoF), scalability, etc. \cite{dMDS}. The standalone MDS necessitates a failover mechanism, otherwise, failure or faulty MDS lead entire cluster down. Furthermore, the consistency policy depends on the designer whether to reflect an immediate effect on modification in standby MDS or after a few seconds. In HDFS \cite{HDFS15}, the secondary NameNode take snapshots which lead to loose consistency of metadata. Furthermore, standalone MDS has to flush the metadata into permanent storage media periodically or on an update. The standalone MDS uses either tree or hash. The hash-based is faster in look-up while tree-based is good in caching. But, hashed-based approach destroys the hierarchical file/directory structure.

\subsubsection{Clustered MDS}
The clustered MDS (cMDS) is a set of MDS that forms a cluster to serve a very large set of metadata. Figure \ref{cMDS} illustrates the architecture of clustered MDS. cMDS is able to serve millions of metadata operations per second. Unlike standalone MDS, the cMDS is free from the bottleneck, fault-tolerance, and small-file problem. The cMDS overcome those disadvantages of standalone MDS. The clustered MDS categorized into two prime categories, namely, distributed MDS (dMDS) and parallel MDS (pMDS). Table \ref{differencetab1} lists the difference between dMDS and pMDS.

\begin{table}[!ht]
    \caption{\textbf{Comparison between Distributed MDS and Parallel MDS}}
    \centering
    \begin{tabular}{|p{5cm}|p{2cm}|p{2cm}|}
    \hline
    \centering \textbf{Parameter} &  \textbf{Distributed MDS} & \textbf{Parallel MDS}\\ \hline \hline
    No of systems & Many & Many \\ \hline
    Bottleneck Issue & No & No \\ \hline
    Scalable & Yes & No \\ \hline
    Flexibility (Changing processor) & Yes & No \\ \hline
    Single point of failure & No & Yes \\ \hline
    Reliability & Yes & No \\ \hline
    Latency Issue & Yes & No \\ \hline
    Hotspot & Yes & NA \\ \hline
    Small File Problem & Yes & Yes \\ \hline
    Rate of parallelism & High & High \\ \hline
    
    \end{tabular}
    \label{differencetab1}
\end{table}

\begin{description}
\item{Distributed MDS (dMDS):}
In dMDS, the metadata is stored in a set of systems forming a cluster. Each system act as an individual system with its own memory space and computation, but communicate with each other periodically as stated in Proposition \ref{prop1}. It is an open challenge to design dMDS. There are diverse parameters to be considered while designing dMDS. The parameters of designing dMDS are a) lowering the communication overhead, b) \textit{ensuring high scalability}, c) \textit{ensuring disaster recoverable}, d) \textit{no hotspot}, e) \textit{lowering the latency}, f) \textit{no down-time}, g) \textit{fine-grained fault tolerance}, g) \textit{improving cache performance}, h) \textit{good load balancing}, and i) \textit{high throughput}.  These parameters are unable to obviate during designing of the dMDS. Undoubtedly, dMDS is highly scalable. On the contrary, the dMDS often face the problem of throughput. Deployed many resources with low utilization rate becomes a barrier in revenue. A dMDS cluster with very low utilization rate does not make any sense. 

\begin{proposition}
\label{prop1}
At least, the dMDS confront the problem of synchronization and network access on an update. It is impossible to obviate these problems. \cite{dMDS}
\end{proposition}

Additionally, the performance of all MDS and loads are not same. The heterogeneity is a barrier in dMDS. The dMDS is very useful in the metadata-intensive computing. Nowadays, the Big Data paradigm is emerging. The Big Data concerns on very large scale data size, ranging from petabytes to Exabytes. Therefore, the standalone MDS is unable to cope with Big Data, and becomes obsolete. The dMDS is a platform to perform metadata-intensive computing. The industries have enough experience in slowing down with $BTree~ or~B^+Tree$ file system in large scale.

\item{Parallel MDS: }
Parallel MDS (pMDS) is similar to dMDS except for the underlying architecture. Unlike dMDS, the pMDS shares memory spaces. The pMDS is deployed in high performance computing (HPC) environment. The pMDS uses fiber optic channel to communicate. In contrast, the dMDS use low-cost commodity hardware. Therefore, the pMDS does not worry about latency and network traffic. The LustreFS \cite{LustreFS15} is most popular file system for HPC system. The HPC system is constructed with Infiniband network, which causes different assumption, but it is very expensive in terms of money. LustreFS designs 1+1 MDS, i.e., every MDS has a backup node for failover. The one MDS has a very high throughput, while other MDS sleeps and its throughput is none for a lifetime. But, the backup MDS can be utilized in load balancing. This kind of shadowing techniques is always not a good design technique. 
\end{description}

\subsection{Data Server}
The data server is the actual location of the data blocks. The responsibility of data server is to provide data to client. For example, Sector \cite{GU} proposed a data storage in optical networks. It is helpful for transmitting scientific datasets securely using s simple API. The data can be replicated/de-duplicated/duplicated in the clustered storage system depending on the requirements. For instance, the HDFS uses the default replication factor as 3 \cite{HDFS15}. On the contrary, the Erasure-coding is used to reduced the storage space requirements in HDFS \cite{EC}. Instead, the replications or duplication system enhances the parallelism as well as assures fault-tolerance, but waste more spaces. In a data server, the data are split into small chunks to achieve high parallelism and fine-grained fault-tolerance. However, most of the data are not accessed for long times, even years; then the replication factor for those data can be reduced. On the contrary, some data are accessed very frequently and it's access rate is exponential. In the context, the data server should replicate the data more than three server to mitigate the load on a few servers.

\section{Methods used in designing MDS}
\label{method}

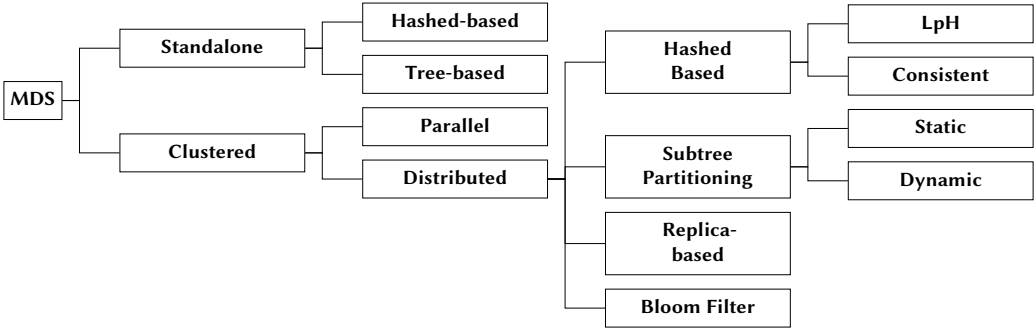
\begin{figure}[ht]
\centering
\resizebox{\linewidth}{!}{
\newcolumntype{C}[1]{>{\centering}p{#1}}
\begin{forest}
 for tree={
  		if level=0{align=center}{
    		align={@{}C{30mm}@{}},
  		},
  		grow=east,
  		draw,
  		font=\sffamily\bfseries,
  		edge path={
    		\noexpand\path [draw, \forestoption{edge}] (!u.parent anchor) -- +(3mm,0) |- (.child anchor)\forestoption{edge label};
  		},
  		parent anchor=east,
  		child anchor=west,
  		l sep=10mm,
	}
  [MDS
  [Clustered 
        [Distributed      
      [Bloom Filter ]
      [Replica-\\based]
      [Subtree\\Partitioning
        [Dynamic]
        [Static]
      ]
      [Hashed\\Based
        [Consistent]
        [LpH]
      ]
    ]
    [Parallel]
    ]
    [Standalone
    	[Tree-based]
        [Hashed-based]
    ]
  ]
\end{forest}
}
\caption{\textbf{Method used to design MDS}}
\label{taxo2}
\end{figure}

\begin{table}[!ht]
    \centering
    \caption{\textbf{Comparison of various methods used in designing Distributed MDS}}   
    \begin{tabular}{|p{4cm}p{1.5cm}p{1.5cm}p{1.5cm}p{1.5cm}|}
   \hline 
     \textbf{Parameters} & \textbf{Hash-based} & \textbf{Replica-based} & \textbf{Tree-based} & \textbf{Bloom Filter}\\ \hline \hline
     
    Standalone or Clustered & Both & Clustered & Both & Clustered\\
    Cache performance & Low & Medium & High & Medium \\
    Hotspot issue & Low & Medium & High & Medium \\
    Load Balancing Issue & Low & Medium & High & Low \\
    Lookup Speed & High & Medium & Low & High \\
    Read performance & High & Low & Low & High \\
    Write performance & Medium & High & High & Medium \\
    Re-adjustment cost & High & Medium & Medium & High \\

    \hline
    \end{tabular}
    \label{tab_dMDS}
\end{table}

A MDS is classified into standalone and clustered MDS. Standalone MDS is further classified into hashed-based and tree-based. Clustered MDS is classified into parallel and distributed MDS. Distributed MDS is further classified into four types, namely, hashed-based, subtree partitioning, replica-based and Bloom Filter. Figure \ref{taxo2} illustrates the classification of MDS. Moreover, Table \ref{tab_dMDS} presents comparison between these MDS based on some parameters. In this section, these MDS are discussed in detail.

\subsection{Hash-based MDS}
Hash-based MDS hashes the file paths for indexing the data and returns the location of the files upon query. It has dozens of advantages, for instance, good load balancing, fast look up, and easy to design cMDS. The hash-based MDS removes the hotspot problems, but, it delivers a very poor cache performance as stated in Trade-off \ref{trof1}, because, the data are placed scatteringly throughout the MDS cluster. The key issue of hash-based MDS is renaming and moving a file, and few such operations slow down the entire cluster as addressed in Dev et al. \cite{DR16}, Xu et al. \cite{DP13}, and  Xiao et al. \cite{SF15}. Nevertheless, the renaming and moving a file occurs rarely. 

\begin{trade-off}\label{trof1}
The Hash-based cMDS provides the best load-balancing capability with the worst locality of references.\cite{dMDS}
\end{trade-off}

The hash-based MDS suffers from cache performance and re-adjustment problems. This issue is resolved by consistent hashing \cite{CH97,DR16} and locality preserving hashing \cite{LpH94,DR16}. The two methods are described as follows:

\begin{description}
\item{Consistent Hashing: }
The consistent hashing \cite{CH97} maps the object into a slot, resize the hash table with a minimal number of movements of keys, which guarantees that there is always consistency even after the redistribution of the hash keys. Besides, the result of consistent hashing does not change under any circumstances, when the result has been used by another function. Consistent hashing is used to eliminate the hotspot problem.

\item{Locality Preserving Hashing (LpH): }
The Locality Preserving Hashing (LpH) \cite{LpH94} is the solution to the poor locality behavior in the cache memory. The LpH does not directly hash, rather, it examines the locality of reference in the cache memory. The LpH does not place data scatteringly. LpH gives the hit performances of cache memory, similar to tree partitioning.

\end{description}

\subsection{Replica-based MDS}
The replica based cMDS replicates the metadata in other MDS, but, it gives the high latency due to network accesses and disk/SSD reads. The cMDS supposed to store all its metadata exclusively in RAM, contrasting to CalvinFS \cite{CF15}, Wang et al. \cite{MR09}. MRFS \cite{Yu14} designs a replicated MDS to deal with the frequently queried metadata, and these metadata are replicated in other MDS. 

\subsection{Tree-based MDS}
The subtree partitioning exhibits \cite{SF15,IF14} the best cache performance, but it exhibits hotspot issue. An MDS can have a huge metadata request, while rest MDS are idle. This phenomenon may cause serious damages to the file systems, and it is not once in a blue moon \cite{dMDS}. Particularly, a file becomes popular for a few hours or a day and a single MDS has to respond millions of metadata requests. Consequently, the MDS becomes a bottleneck. Apparently, the rest of the MDS are idle. Also, as Trade-off \ref{trof2} defines the trade-off of tree-partitioning between cache performance and load-balancing.

\begin{trade-off}\label{trof2}
The tree partitioning exhibits the best cache performance with least load-balancing capability.'' \cite{dMDS}
\end{trade-off}
The metadata splitting is a primary issue in tree-based MDS. The splitting is solved by two methods, which are given below-
\begin{description}
\item{Static Subtree Partitioning:} 
The static subtree partitioning is method to distribute the metadata among the MDS and it never re-balances the loads of a particular MDS after the subtree partitioning. Once partitioned the metadata, it remains same, and fixed the MDS servers for serving the metadata. The static subtree partitioning provides an excellent cache performance, for instance, AFS, NFS, Coda, and Sprite. The metadata can be foisted somewhere, but, once allocated in an MDS, never changes in its entire lifetime \cite{dMDS}.

\item{Dynamic Subtree Partitioning:}
The dynamic subtree partitioning balances the load dynamically to avoid the bottleneck of MDS. Its cache performance is similar to static subtree partitioning. However, the load can be re-adjusted dynamically upon bottleneck of a particular MDS.
\end{description}

\subsection{Bloom Filter}
Bloom filter is one of the most famous data structure for approximate membership query and it is introduced by Burton H. Bloom in 1970 \cite{Bloom70}, which has met wider application areas. It is a probabilistic data structure to test whether a given item belongs to a set or not. The Bloom filter has two possibilities while testing the membership of an element in a set, namely, ``possibly in the set'' or ``definitely not in the set''. Thus, ``definitely not in the set'' is very useful in testing a membership of an element in a set. It provides a quick response to a query. However, the Bloom filter may return false positives, but impossible to return a false negative in standard Bloom filter. 

\begin{proposition}\label{prop2}
Bloom filter is used as an additional technique to boost up the performance of an MDS.
\end{proposition}

There are numerous variants of Bloom filters. Almeida et al. \cite{Almeida} implements a scalable Bloom filter where the Bloom filter re-adjust dynamically. In addition, Fan et al. \cite{Fan} implements better look up performance than Bloom filter. This work can be used to filter large-scale membership test, like metadata. BlooM Filter is currently used to handle huge sized IoT data \cite{SINGH}. Bloom filter is a useful data structure to enhance metadata performance as stated in Proposition \ref{prop2}.


\section{MDS designs}
\label{mmds}

\begin{figure}[!ht]
\centering
\resizebox{\linewidth}{!}{
\newcolumntype{C}[1]{>{\centering}p{#1}}
\begin{forest}
 for tree={
    align=center,
    parent anchor=south,
    child anchor=north,
    font=\sffamily,
    edge={thick, -{Stealth[]}},
    l sep+=10pt,
    edge path={
      \noexpand\path [draw, \forestoption{edge}] (!u.parent anchor) -- +(0,-10pt) -| (.child anchor)\forestoption{edge label};
    },
    if level=0{
      inner xsep=0pt,
      tikz={\draw [thick] (.south east) -- (.south west);}
    }{}
  }
  [\large MDS Design
  	[ \large Tree-based 
    	[LocoFS \\SoMeta \\HopsFS \\Virtual MDS \\Replichard \\PPFS \\MAMS \\CephFS \\Partitioner \\ShardFS \\IndexFS \\CEFLS]
    ]
    [ \large Hash-based 
    	[Giraffa \\AbFS \\DROP \\Dr. Hadoop]
    ]
    [ \large Load-aware [HWM \\Adaptive Metadata Rebalancing]
    ]
    [ \large Cache-aware [C$^2$\\DDCache\\DEAM]
    ]
    [ \large Geo-aware [SATURN\\MRFS\\CalvinFS]
    ]
    [\large Client-funded [DeltaFS \\ BatchFS]
    ]
    [\large Bloom Filter [MaaS\\ MBFS \\G-HBA \\HBA]
    ]    
  ]
\end{forest}
}
\caption{\textbf{State-of-the-art MDS designs}}
\label{taxo3}
\end{figure}
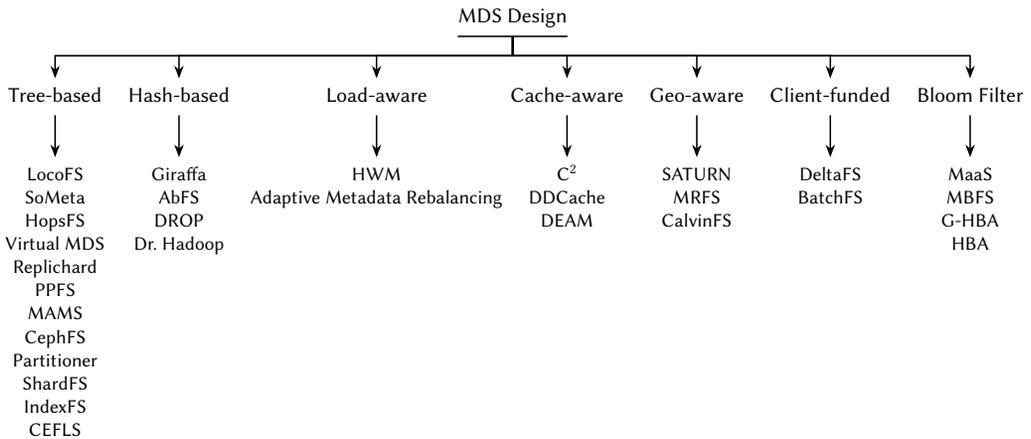

MDS designs are categorized into seven key categories, namely, Bloom filter, client-funded MDS, geo-aware MDS, cache-aware MDS, load-aware MDS, hash-based MDS, and tree-based MDS. Figure \ref{taxo3} has listed the proposed MDS based on the MDS designs. Moreover, Table \ref{tab1} compares the most famous MDS for file systems. 

\subsection{Bloom Filter}
In Bloom filter based MDS, the MDS uses Bloom filter for its metadata management. Many variants of Bloom filters are proposed. Some Bloom filter based MDS are discussed in this section.

Zhu et al. proposed Hierarchical Bloom filter Arrays (HBA) \cite{Zhu04,Zhu08}, a MDS that is based on Bloom filter. It uses Bloom filter hierarchically. At the first level, a small Bloom filter is used to capture the MDS information. Second level has a pure Bloom array (PBA). The MDS that creates the Bloom filter is called home MDS. Home MDS prepares the Bloom filter by inserting all metadata present in that MDS. Later, the Bloom filter sends to all other MDS. All MDS maintains Bloom filter from all other MDS, hence, it has to maintain Bloom filter arrays. When the client makes a request a MDS is chosen randomly and the metadata is searched. The MDS query all Bloom filters. When a Bloom filter return true, then the metadata is sent to the client. When no Bloom filter returns true or multiple Bloom filter returns true, then the false response is sent to the client. Both Bloom filters are replicated in all MDS. It speedup the local query operation. HBA has low accuracy, but high memory efficiency. 

Hua et al. \cite{Hua11} implements Group-based Hierarchical Bloom filter Array (G-HBA) by extending HBA \cite{Zhu04,Zhu08}. The G-HBA is a Bloom filter array to correctly route the MDS. There is a set of MDS and a client can query to any MDS about metadata. The G-HBA is comprised of many MDS to scale nearly Exabytes. The Bloom filter array is used to route directly to the correct MDS among the cluster of MDS.  The G-HBA maintains a replica of $\frac{N-M'}{M'}$ where $N$ is a total number of MDS, $M'$ is the number of MDS in groups. G-HBA does not use a hash function, because it is very costly to implement and maintain a dynamic hash-based system. 

MBFS \cite{Huo15} is a lightweight metadata query technique which uses a tiny amount of memory. The MBFS runs concurrently in every MDS to improve metadata queries. MBFS creates a Bloom filter for each metadata attribute, then they are combined to form multi-dimensional Bloom filters (MDBF). To achieve faster searching, MBFS trims directory subtree at upper hierarchy using the breadth first search algorithm. Furthermore, the MBFS uses MapReduce paradigm for searching a key. Conversely, the MapReduce programming is based on a file system. The metadata is stored in files and processed using $map$ and $reduce$. But, the MDS with an in-memory design is faster than file system based. However, the in-memory design lacks scalability. The MBFS does not support the delete operation, because MDBF uses a standard Bloom filter which does not support delete operation. Instead, counting Bloom filter can be used to support delete operation.

MaaS \cite{Anitha15} performs fast retrieval of cloud data. It is based on Bloom filters. It introduces new Cloud Bloom filter array (CBF) which is comprised of Global Bloom filter (GBF), Local Bloom filter (LBF) and Bloomier Matrix Filter (BMF). The GBF is responsible for the creation and placement of metadata. The LBF is responsible for metadata updates. And finally, the BMF is responsible for analyzing the metadata. This approach has significant advantages of retrieval of data and metadata in a cloud environment.

\subsection{Client-funded MDS}
In client-funded MDS, the client saves the metadata for future use. When the client request a metadata, the MDS sends the metadata to the client. The client stores the metadata in its cache to use in the future. The metadata is periodically refreshed. It reduces network traffics.
\begin{landscape}
\begin{table}[!ht]
\centering
\caption{\textbf{Comparison of available  MDS design. `-' defines unknown in the table. Source \cite{MDS}}}
\resizebox{\linewidth}{!}{\begin{tabular}{|p{1cm}|p{1cm}|p{1cm}|p{1cm}|p{1.2cm}|p{1cm}|p{1cm}|p{1cm}|p{1cm}|p{1cm}|p{1cm}|p{0.7cm}|p{1cm}|p{1cm}|p{1cm}|p{1cm}|p{1cm}|p{1cm}|p{1cm}|p{0.8cm}|p{0.8cm}|p{0.5cm}|} 
\hline
  \centering \rotatebox{270}{\textbf{Name}}  &  \centering \rotatebox{270}{\textbf{Scalability}} & \centering \rotatebox{270}{\textbf{Hot-standby}} &  \centering \rotatebox{270}{\textbf{Load Balancing}} &  \centering \rotatebox{270}{\textbf{Cache Performance}} &  \centering \rotatebox{270}{\textbf{Latency}} &  \centering \rotatebox{270}{\textbf{Number of RPC}} &  \centering \rotatebox{270}{\textbf{Hotspot Issue}} & \centering \rotatebox{270}{\textbf{Hiccup Issue}} &  \centering \rotatebox{270}{\textbf{Write Throughput}} &  \centering \rotatebox{270}{\textbf{Read Throughput}} &  \centering \rotatebox{270}{\textbf{SPoF}} &  \centering \rotatebox{270}{\textbf{High Availability} } & \centering \rotatebox{270}{\textbf{Renaming Overhead}} &  \centering \rotatebox{270}{\textbf{Small File Problem}} &  \centering \rotatebox{270}{\textbf{Communication Cost}} &  \centering \rotatebox{270}{\textbf{Distributed Lock on Metadata}} &  \centering \rotatebox{270}{\textbf{Up-time}} & \centering \rotatebox{270}{\textbf{Posix-Compliant}}  &  \centering \rotatebox{270}{\textbf{Disaster recovery}} & \rotatebox{270}{\textbf{Geo-aware}} \\ \hline 

  HDFS & Limited & No & NA & Excellent & Neglig- ible & NA & Yes & Yes & Medium & Medium & Yes & Depend- ent & No & Yes & Neglig- ible & No & Depend- ent & No & No & No \\ \hline
  
  QFS & Limited & No & NA & Excellent & Neglig- ible & NA & Yes & Yes & High & High & Yes & Depend- ent & No & Yes & Neglig- ible & No & Depend- ent & No & No & No \\ \hline
  
  GFS & Limited & No & NA & Excellent & Neglig- ible & NA & Yes & Yes & - & - & Yes & Depend- ent & No & Yes & Neglig- ible & No & Depend- ent & No & No & No \\ \hline
  
  CEFLS & Limited & No & No & Excellent & High & High & Yes & Yes & Medium & Medium & No & - & Less & No & High & Yes & - & No & No & No\\ \hline
  
  DROP & Highly Scalable & No & Excellent & Good & Medium & Medium & No & May be & Good & Good & No & Yes & Yes & No & Low & Yes & - & No & No & No \\ \hline
  
  IndexFS & Limited & No & Bad & Very Good & Very high & Very high & Yes & Yes & - & - & No & Yes & Yes & Can be & High & Yes & - & No & No & No \\ \hline
  
  ShardFS & Limited & Yes & Yes & Excellent & High & Very High & Yes & Yes & - & - & No & Yes & Very High & Yes & High & Yes & - & No & No & No\\ \hline
  
  BatchFS & Scalable & Yes & Yes & Excellent & High & Low & Yes & Yes & - & -& No & Yes & Yes  & Yes & High & Yes & - & Yes & No & No   \\ \hline
  
  CalvinFS & Scalable & No & Yes & Excellent & Depends  & Low & Yes & Yes & Good & Good& Yes & Yes & High & Yes & High  & Yes & - & Yes & Yes & Yes\\ \hline
  
  Dr. Hadoop & Infinite & Yes & Good & Good & Low & Medium & No & Auto- matic & Good & Good & No & Yes & High & No & Medium & Yes & 99.99\% & No & No & No\\ \hline
  
  DMoose- FS & Limited & Yes & Minimal & Good & Low & Low & Yes & No & - & -& No & Yes & Yes & Can be & High & - & - & No & No & No\\ \hline
  
  CephFS & Infinite & No & Yes & Very Good & Low & Medium & No & Yes & - & -& No & Yes & Yes & No & High & - & - & Yes & No & No  \\
  \hline
\end{tabular}
}
\label{tab1}
\end{table}
\end{landscape}

BatchFS \cite{BFS14} is a client-driven MDS. BatchFS cluster lies above scalable storage architecture. It has three types of servers, namely, primary, auxiliary, and private. The primary MDS are the dedicated servers. It manages the global image of the file system. Each server has a non-overlapping set of directory partition. Interactive clients continuously communicate with private MDS to maintain a global namespace. Auxiliary and private MDS are client-funded MDS. They run on demand, temporarily using client resources. Using these MDSs, clients modifies their private namespace locally. Auxiliary MDS is daemon processes that merges client-side namespace into global file image. It uses SSTables to store the metadata. The clients of BatchFS synchronize when the clients want to merge the modified namespace in MDS. This approach allows direct access to the data without interacting with the MDS. However, thousands of client modifying same metadata creates a bottleneck to the MDS. The write intensive file system cannot afford the cost of the bulk amount of synchronization. In addition, the security system is another issue with client-driven MDS, which is more vulnerable to DDOS and Spoofing. This security system also gives additional cost in terms of latency, bandwidth and money.
 
The DeltaFS \cite{DeltaFS15} is an MDS that does not support dedicated MDS or global file system namespace. DeltaFS lies above the data storage system and manages all metadata operations. For metadata service, DeltaFS application links to a user library and initialize a set of auxiliary MDS that helps to interact with namespace partitioning. The batch application takes appropriate namespace present in public registry and manage their own namespace without coordinating with other applications. In DeltaFS, the namespace is stored in SSTables. SSTables are arranged using LSM-Trees and namespace are managed by LevelDB. DeltaFS uses two namespaces partitioning strategies, type-I (overlapping sub-namespaces) and type-II (partitioned sub-namespaces). Type -I consist of a set of non-communicating (but related) processes which execute metadata operation using own embedded private metadata service. Type-II consists of a set of processes that require metadata produced by another application in the middle of the run. DeltaFS has good performance due to concurrent global compaction of SSTable. This system deals with HDD and does not have any failover mechanism.


\subsection{Geo-aware MDS}
In Geo-aware MDS, the systems are present in different geographical locations. Due to long distance between the system consistency, synchronization, network issues and other issues need to be addressed. 

CalvinFS is a metadata management layer which partition the file system horizontally and replicates them in a shared-nothing cluster across different geographic regions. It has three main components, namely, a transaction request log, a storage layer, and scheduling layer. The log stores global sequence of totally ordered transaction requests. The scheduling layer is responsible to execute logged transaction requests in a serial manner exactly in the order they appear in the log. The log is implemented using three components, namely, front-end servers, block store and meta-log server. The client request is sent to front-end server. Front-end server appends the request in the log. Front-end server batches it with other incoming request and write them to block-store. Then the front-end server sends the block id to the meta-log server. The CalvinFS uses Optimistic Lock Location Prediction (OLLP) to determine read and write sets of a compound transaction. The deterministic locking is deadlock free, serializable, and lack a distributed commit protocol for distributed transactions. Hence, it reduces latency and improves scalability of the system. Further, the metadata is hashed based on full file path which gives better load balancing and simplifies data placement tracking. Also, under heavy load, occasional background task (e.g. garbage collection) causes stalling of many concurrent read requests, occasional latency spikes and some failed reads that need to be re-executed.

The Metadata Replication File System (MRFS) \cite{Yu14} is an MDS having efficient hierarchical distributed metadata management. The MBFS is extended from MooseFS’s MDS module. It has four components, namely, client, MDS, NameSpace Server (NS), and Chunk Server (CS). The MDS are distributed in different geographical locations. The NS is a single server,  which has the following responsibilities (a) maintains the global namespace and the whole file system (b) a consistent namespace in RAM, and (c) stores the mapping between metadata entry to the primary MDS. The CS provides storage for the data files. The MRFS client is built on FUSE. First, the client establishes a long lived TCP connection and register with primary MDS. The primary MDS handles most of the operations. Each MDS forwards all namespaces related operation to the NS. When a client makes a request, the primary MDS search locally. If found, then it returns directly to the client. Otherwise, the request is forwarded to the NS. The NS returns the locations of the metadata if the path exists. The primary MDS connects to the Host MDS (MDS, which stores metadata entry) and request for metadata. Further, the MDS uses a hash table to store metadata. The NS improves efficiency of directory query operation. Furthermore, the NS uses a single thread model to reduce the consistency issue. The primary MDS stores metadata in RAM, which makes the operational speed faster. Also, the MRFS automatically removes stale replicas. It uses a replication mechanism which solves the hotspot problem.

The SATURN is a metadata service that can easily attach to existing geo-replicated data services to effectively ensure causal consistency across geo-locations. The metadata is divided into small chunks of constant size, called labels. When update request is received from the client, the datacenter generates labels. Then, the payload is attached to labels based on the order that respect causality. Next the labels are passed to the SATURN for bulk transfer. The SATURN then delivers them to the appropriate datacenter in causal order. After that, the datacenter applies the update locally. The SATURN keeps the metadata size constant and smaller irrespective of number of clients, MDSs, partitions, and location to improve efficiency. The timestamp of causality is used to order the labels globally. It makes them robust to failures. But, the SATURN cannot achieve optimal label propagation latency for every pair of datacenters.

\subsection{Cache-aware MDS}
The Decoupled, expressive, area-efficient metadata (DEAM) \cite{Liu14} is a metadata cache framework to save storage of metadata, to reduce latency of data access, and coherence activities. It has three types of entities in the metadata cache, namely, home, keeper and sharer. The home entries stores keeper location. The entries of keeper and sharer manage the access and coherence tracking. In DEAM, Home Metadata Cache (HMC) is implemented for home metadata entries (HME), the coherence metadata cache (CMC) is implemented for keeper metadata entries (KME), and the sharer metadata entries (SME). Initially, the level-one data-cache (DL1) copy is created by the requester. The HME record DL1 location as keeper. When the private level-one data-cache (DL2) copy is evicted, it is stored in DL2 slice as a victim. Cache misses are handled in this hierarchy CMC, local DL2, and finally in home node. For delegation of keeper, in case of read-write, data keeper is selected among the sharer based on dynamic access pattern. In general, dominant writer is chosen as the keeper to allow two-hop invalidation. And private data can avoid coherence tracking to save storage. For read-write data, keeper use full bit vector to know global sharer and sharer use pointers to store keeper ID. For read-only data, keeper uses a short vector to know sharer and the sharer use pointer to know a nearby sharer. HME is used to recognize an access pattern. The DEAM’s flexible unified structure reduces conflicts of data. Furthermore, the metadata cache improves stream prefetching. When a block is delegated it is stored in DL2 of home node which improves the usage of DL2 space and reduces off-chip latency. Due to better data cache utility. Moreover, it reduces traffic and energy consumption. DEAM makes a trade-off between the latency and capacity.

DDCache (Decoupled, Delegate Data \& Cache) \cite{Hossain09} is a cache coherence protocol that decouples metadata and data in a cache set. A tile in the core multiprocessor (CMP) contains a processor core, private level-one (L1) cache, and a portion of level-two (L2) cache. L1 cache also contains sharer list. And, L2 cache is globally shared and it is organized to decouple data and metadata. An address based destination predictor is used to predict the location of owner of the data or closest sharer. This predictor table has a cache-like structure which have a valid bit, a tag, and destination node’s processor ID. L1 cache miss is handled in this hierarchy, predictor table then home node. Moreover, DDCache delegates decide to find the access pattern. Furthermore, the Coherence protocol to access pattern is used to avoid unwanted communication and indirection during data sharing. It achieves faster coherence, and reduces traffic on both interconnection in on-chip and off-chip. Decoupling of data and metadata helps in avoiding unwanted data replication. Also, DDcache improves execution speed which reduces the coherence latency, which also reduces the energy consumption. 

The Cloud Cache (C$^2$) \cite{Xu15} is an adaptive load balancing scheme for MDS cluster. It adopts two caching scheme to increase the number of replicas. It helps to reduce the workload of MDSs. These schemes adapt to changing workload in the EB-Scale storage system. In caching scheme, two adaptive cache diffusion techniques, namely, load stealing and load shedding is implemented. In load stealing, an underloaded node search for overloaded nodes. If found, a replica of the metadata is created in that node and a redirection pointer is created to the underloaded node. In load shedding, an overloaded node search and gives some loads to one or more underloaded node. In replication scheme, two replica diffusion schemes are used, namely, directory based and chain based. In directory based scheme, when the number of requests for a metadata in a node exceeds its threshold, the node creates a directory for that metadata. In chain based scheme, when the number of requests for a metadata in a node exceeds its threshold the node create a replica in the hop closer to the source of last request. The C$^2$ have excellent scalability. Furthermore, metadata pointers reduces metadata migration overhead. But it temporarily hurts metadata locality when balancing the load. Also, the directory based scheme has three advantages compared to chain based. Advantages are faster replica transmission speed, higher query parallelism, and good load balancing.

\subsection{Load-aware MDS}
Cha et al. \cite{Cha2017} proposed a metadata model that performs adaptive metadata rebalancing with minimum metadata operation failure. Adaptive metadata rebalance architecture consists of Migrator, Monitor, and Controller. The Migrator relocates metadata to another MDS. The Monitor stores the status of migration and load information of the cluster. The Controller decides to activate or deactivate Migrator in MDS based on information provided by the Monitor. The Controller selects a candidate directory for metadata rebalancing using some selection conditions. Then it instructs source MDS to migrate metadata to target MDS. Also, the Migrator notifies to update rebalance array in the directory. This model reduces interference between metadata rebalance and normal metadata operation to minimize. In addition, it can decide the feasibility of metadata rebalancing. Again, this approach maintains feasible and stable metadata rebalancing under extreme conditions.

Hybrid Workload Migration (HWM) \cite{HWM} is a mechanism to efficiently migrate the workload of the MDS. The mechanism contains three parts, namely, metadata service migration, state service migration and request redirection. The metadata service migration transfers the metadata volume. The state service migration reconstructs all of the state-sets of the metadata volume. The request redirection classifies the request operations and decide which MDS should process the request. The HWM reduces metadata and state access latency. Furthermore, it uses compound and multithreading technique to increase migration process.

\subsection{Hashed-based MDS}
The Dr. Hadoop \cite{DR16,nayak1} is designed to achieve the infinite scalability of metadata. It is based on the Doubly Circular Linked List. Once MDS is full, a new node is inserted into the MDS cluster. The load balancing is done using consistent hash-based metadata. The Dr. Hadoop provides bundles of features, namely- Infinite Scalability, Hot-Standby, 99.99\% uptime, i.e., no downtime, LpH provides cache performance, individual MDS has less load, designed using minimal number of servers, no single point of failure, and no bottleneck on MDS. Some drawbacks are, namely- communication overhead, problems of renaming and moving files. Coordination is required in every update, and insertion of new MDS. The initial configuration of Dr. Hadoop is three nodes if the size of metadata is less. The node of Dr. Hadoop may grow to $n$ node, depending upon the size of metadata. A new node is inserted only when the metadata size exceeds the RAM size of any given node. The insertion is performed between two nodes same as a circular doubly linked list. The left and right side of the node is defined logically by assigning the IP address to them. Therefore, one copy of metadata has stored in both, right and left node of a given node. These left and right nodes act as hot-standby of the node. When this node fails, one of the node act as a main node (either left node or right node). The left node has also left and right hot-standby node. Similarly, the right node also have left and right hot-standby node, and so on. Thus, circular doubly linked list is maintained.

The Dynamic Ring Online Partitioning (DROP) \cite{DP13} is a ring based MDS. The DROP uses locality preserving hashing (LpH) \cite{LpH94} and chord protocol \cite{Chord}. DROP creates virtual nodes and actual nodes for MDS. The storage manager of MDS manages storage system having SSD/NVM-base key-value under three components, i.e. replication engine, failover policy and lookup cache. The replication engine is used for reading and writing metadata. The lookup cache is used to find hotspots and replicate them. The DROP uses Histogram based load balancing (HDLB) technique for load balancing. The DROP uses LpH to improves namespace locality, and to eliminate hierarchical directory traversal overhead. It locates appropriate MDS for a metadata request using a message to a single MDS. The load balancing is done in parallel. The virtual nodes create a network traffic problem as they keep sending messages to know whether an MDS is alive or not. If not, then the dead MDS is replaced by new nodes. And, as each real MDS act as many virtual nodes, and therefore, the network traffic is multiplied. The metadata publishing using LpH creates two problems. First identical content files, but with different paths can be stored on different MDS because of their different keys. Second, if a replica group fails, then most of the metadata are not available to few clients.

The AbFS \cite{Diaz13} is a distributed file system which deploys hash/table based mapping approach to reduce inode lookup time. It helps to share the inexpensive devices attached to the commodity computers efficiently in the cluster. The AbFS uses several structures for management of metadata, namely, volume table, delegation table, inode/dentry structure, and access table. The hash function or delegation table (hash table) divides the metadata into segments and assigns them to the volumes. The volumes and delegation tables are replicated in all nodes. AbFS employ both client and MDS cache to increase performance. It also takes advantage of the LINUX metadata caches to avoid the requirement of additional layers.

The Giraffa is a distributed, highly scalable file system that plans to scale both metadata and data storage. All data are stored in DataNodes. DataNodes stores the data in Hbase table. The Hbase supports automatic dynamic partitions of the namespace table. The file's list of blocks and their location is given for reading a file, then data can be read from respective DataNodes. The BlockManger allocates block and client to write data in the DataNode in a file. The BlockManager is responsible for the allocation of the new block, block replication, deletion scheduling, data management, and  management of the flat namespace of the blocks. The Giraffa needs to address problems like distributed block management, and Hbase scalability.

\subsection{Tree-based MDS}
The Cost-Effective File Lookup Service (CEFLS) \cite{CEFLS12} is based on directory partition method to achieve higher cache performance. The CEFLS consist of three main parts, namely, partitioned cache, partition manager and data persistence layer. The partitioned cache adapts efficient in-memory data structures to represent each directory partition (DP) metadata. The partition manager is solely responsible for partitioning the directory tree for distributing the metadata. And, the data persistence layer is based on disk and memory. In this design, the clients randomly connect to any MDS and the MDS redirect it to the correct MDS. 

The IndexFS \cite{IF14} is a middleware MDS design that is implemented on existing file systems like PVFS, Luster and HDFS to improve their efficiency. The metadata or data request for file size less than 64KB is handled by the IndexFS and for large file size it is redirected to clustered file system. The IndexFS puts the newly created directory in random server and splits the directories based on GIGA+ \cite{GIGA+}. It utilizes client caching of directory entries to mitigate hotspots. The metadata is stored in SSTable and these are arranged using LSM tree. The IndexFS applies a second LSM table, called column-style that only stores final pathname component string, permission and a pointer. The standby servers are used for fault tolerance. Use of column-style prevents compaction of full MDS. The GIGA+ caches the mapping of directory partition, which eliminate all RPC roundtrips needed to find the correct MDS. IndexFS client store frequently used metadata to enhance cache performance. The small files are stored in the metadata entry, which increases the data volume when processed by each compaction. 

The ShardFS \cite{SF15} deploys optimistic and pessimistic concurrency control mechanism. The ShardFS  puts entire directories to all servers, such that any server can serve director look up. The problem arises only if the directory size exceeds the RAM size. This technique of duplicating directory lookup helps to reduce the number of RPC for all operations. In ShardFS directory metadata updating is slow, so, its scalability is limited. But, it is very suitable for a small scale file system, because ShardFS provide perfect load balancing and reduced RPC.


The Partitioner \cite{PT14} is a coordinator of collected NameNodes. The Partitioner accept client request and forward it to desired NameNode. The Partitioner controls NameNodes when write request arrives from the client. The Partitioner uses a backup NameNode and it is used upon failure of master NameNode. The Partitioner scalability is infinite, due to the commodity of NameNodes. On the contrary, the Partitioner introduced another network layers. This cost is expensive to any other cMDS design in terms of latency, network traffic, network failure, and a probability of network partition from client to partition or Partitioner to NameNodes.


Multiple actives multiple standbys (MAMS) \cite{MAMS} is a metadata service that keeps multiple standby in case of failure in a parallel and distributed system. The MAMS make replica group consist of one active, and two or more standby nodes. MDS can be in three states active, standby or junior. An active MDS is responsible for the client request management and the data storage management. The hot-standby node keeps up-to-date namespace state with active node. The junior MDS is a passive standby server. When an active MDS fails, the standby MDSs elect the active MDS using the Paxos algorithm. The standby MDSs compete to acquire distributed lock which is possessed by active MDS. The MAMS policy has high reliability so metadata services can be provided efficiently. In MAMS hot standby reduces recovery overheads and increases reliability and availability of metadata services. As the standby is hot, the restarting time and reconnection time is reduced. Keeping several backups increases reliability, but also leads to wastage of nodes. MAMS policy reduces performance of some metadata operations to become reliable.

The Post Peta-scale File System (PPFS) \cite{PPFS} is an MDS for post peta-scale system. It is implemented using Post Peta-scale MDS (PPMDS) and post peta-scale object storage server (PPOSS). PPMDS is a scale-out distributed MDS and PPOSS is a distributed object storage server using PPOST as backend storage. PPMDS consist of an MDS, called PPMDS and a client software using FUSE, called PPFUSE. It manages metadata on a key / value store which enables range search based on inode number of the directories. PPOST is an object storage using openNVM. This is an implementation of flash memory to achieve higher Input/Output Operations per second (IOPS). Use of PPMDS gives a high file creation performance and use of PPOST as backend storage increase I/O performance. Network traffic is reduced by the storage server in compute mode. Communication overhead between MDS and storage server is also reduced by bulk creation. In PPOSS, files are not distributed throughout the PPOSS nodes and few clients use PPOSS, so, the block size is made smaller. In PPFS, PPOSS is independent of PPMDS which make the I/O performance of PPOSS scalable.

The Replichard \cite{Replichard} is a metadata service provider which combines two consistency scheme, namely, (a) a strict consistency is implemented for non-idempotent requests and (b) a relaxed consistency is implemented for the idempotent requests, to achieve high throughput. All MDSs are registered in a global registry. When a client makes non-idempotent request for a file, global registry assigns a write-lock to an MDS for that file path. Thus, all writing is performed by its owner MDS. In case of idempotent request, any MDS can respond. Replichard uses two techniques for metadata update, namely, instant broadcast and the subtree-barrier. In instant broadcast, after receiving updated file, the MDSs just replaces the file. In subtree-barrier, entire subtree is packed and broadcasted. A master MDS is elected to reduce network traffic, which receives DataNode heartbeat messages and block report, and broadcast it to all MDSs. The strict consistency is used for write operations and relaxed consistency for reading operations to speed up user applications. But, Replichard does not focus on MDS failure recovery. Further, subtree-barrier technique is not efficient as most of the updates are already transmitted using instant broadcast.

Virtual MDS \cite{vMDS} is proposed to give multi-level consistency and high performance. Multi-level consistency is implemented by combining the features of metadata services and virtual disks without any master MDS. In virtual MDS, instead of data nodes proxy nodes store the metadata. Proxy nodes are located at the creation location of the virtual disks. Because proxy nodes store metadata of virtual disks. Virtual MDS implements object based key-value scheme for metadata storage. No master node, therefore, each virtual disk is owned by a single MDS proxy node. Virtual MDS performs an asynchronous update of metadata. However, this reduces the read operation performance. And, uses eventual consistency to achieve consistency in data storage system. Virtual MDS implements LRU technique for metadata management. However, LRU introduces small overhead. However, the metadata is cached to reduce the overhead. 

HopsFS \cite{hopsfs, hopsfs2} is the replacement of HDFS \cite{HDFS}. It builds on NewSQL database and solves the single-point of failure issue. Moreover, HopsFS has lower delay for concurrent clients and no downtime during failure. HopsFS stores the metadata in a commodity database called Network Database (NDB). The NDB is distributed, in-memory, highly available relational database. NDB helps in failure recovery both in DataNode and cluster level. Moreover, NDB helps in increasing scalability and metadata migration and free-text search. HopFS increase the performance  of file system operations by combining both traditional database techniques and write-ahead caches in a transaction. HopsFS partitions the namespace such that directory’s all immediate descendants are stored in the same database. This improves the directory listing. In addition, HopsFS implements inode hints cache for quick searching of file paths. Hints \cite{hints} is a technique for finding the path component in parallel i.e. batched operations. HopsFS uses ePipe \cite{ePipe} which is a databus. It creates a consistent change stream that is delivered to clients in the correct ordered stream. ePipe provides polyglot storage to MDS. It helps in handling metadata query efficiently. HopsFS implements block reporting \cite{hopsfs1}. Block reporting is a technique of DataNodes sending periodic ground truth report on all file systems block to MDS. This helps in synchronization of the current state of the data blocks. HopsFS also implements an application level distributed locking mechanism. It separates the subtree (on which operation is performed) from the rest of the tree. In addition, large file system operations are partitioned into smaller parts for parallel execution. HopsFS metadata design and partitioning of metadata helps in performing common operations in low cost. Metadata is saved in tables and each row is a single directory inode. Incorrect hints increases network traffic. HopsFS cannot identify the NameNode facing hotspot issue. In case the metadata is completely available in RAM, then HDFS is better than HopsFS. 

SoMeta (Scalable Object-centric METAdata management) \cite{SoMeta} is an MDS in HPC systems. SoMeta is a decentralized MDS. It dynamically partition the flat namespace. It helps in efficient metadata management, searching and update of metadata. It also provides a fault tolerant and light-weight metadata management technique. The fault tolerance technique is a window-based adaptive technique. Moreover, it recovers from failure without data loss. SoMeta consider each metadata as object. Metadata attributes are saved as tags. Tags helps in easy logical group formation. Moreover, tags helps in searching and updating the attributes efficiently without centralized synchronization. SoMeta uses counting Bloom filter \cite{CBF} to speedup the search process for same name metadata. Distributed Hash Table helps in transmitting metadata object to other servers in the cluster. SoMeta increases the efficiency of metadata access by using two level hashing. First level hashing separate the namespace among servers. Second level hashing value is the hash key to insert into the hash table. SoMeta has high performance because it does not maintain metadata replica. It eliminates consistency maintenance overhead. However, it leads to data loss in case of system failure. SoMeta sent the client request to only one server to eliminate consistency overhead. However, the performance of the system is reduced in case of the hotspot.

LocoFS (LOosely-COupled metadata File System) \cite{locofs} is a distributed file system that decouples the dependencies among different metadata types to decrease network delay and increases the efficiency of using key-value store. LocoFS partition the metadata file inode from a directory tree. Metadata files are organized separately to form a flat space. However, the relationship information is kept in the file inode using reverse indexing. Flat space helps in efficient key-value access. LocoFS consists of client, Directory Metadata Server (DMS), File Metadata Server (FMS) and object store. DMS stores the metadata directory. FMS stores metadata of files. LocoFS have single DMS and multiple FMS. Single DMS is capable of handling large number of requests to access the directories. DMS also saves a large number of directories due to flat namespace design. Moreover, single DMS helps in easy verification of Access Control List. However, single DMS leads to many issues such as single-point-of-failure, scalability, bottleneck and so on. To increase scalability and reduce network latency the metadata is stored in the cache of the client system. LocoFS uses multiple FMS and distribute the metadata file using consistent hashing. In LocoFS, the $rmdir$ operation is costly because all its directory entries need to be collected from all servers. LocoFS have a high cache miss ratio for d-inode. With the increase in depth of directory tree the access control list checking takes more time. Hence, the file creation time increases.


\section{Issues in MDS design}
\label{issue}

MDS plays a vital role in the Big Data to enhance the data accesses. MDS has a significant co-relation with Big Data. All functions of file system stops without MDS, which is a cardinal part of the file system (data storage). MDS uses an in-memory computation model. These data are prone to lose because of the volatility nature of RAM. The data become meaningless when there is no metadata. Therefore, the design must take care of every issue pertaining to it. The advantages of cMDS are stated in Proposition \ref{prop3}.

\begin{proposition}\label{prop3}
A cMDS comprise of at least high scalability, durability, fine-grained fault-tolerance, high throughput and high availability.
\end{proposition}


\subsection{Bottleneck}
The standalone MDS clearly shows bottleneck of a file system in millions of metadata request. The metadata resides in RAM. The data are rapidly increasing, and thus, Big Data becomes a revolutionary research area. The growing clients along with rapid data growth is termed as Vendee \cite{RP16}. The Vendee sizes are millions, as a consequence, billions of metadata operations are performed. The massive data cohere with large set of metadata along with billions of metadata operations. Apparently, the huge number of metadata request slow-down the system or become bottleneck. 

\subsection{Small File Problems}
It is a well-known fact that Hadoop NameNode has experienced with excessive metadata generation with very small sized files \cite{Ev,HAR+15,Dev2016}. MDS can be out of RAM space if the file size is very less. A large number of small files can  generate huge metadata size while the actual data size is comparatively low. For instance, a set of file size is less than 10 MB and it can consume entire RAM space of MDS. Many research papers address the small sized file issue, for instance, HAR \cite{HAR15}, and HAR+ \cite{HAR+15}, TyrFS \cite{TyrFS}. The small file problem seriously affects the performance of the file system. 

\subsection{Fault-tolerance and Availability}
The fault-tolerance is the big issue in MDS designing. For instance, an MDS fails, or become faulty, then what are the corrective steps? What is the administrative cost? Is there the hiccup time for MDS? Does the MDS design taken care of single point of failure (SPoF). For instance, Apache Hadoop suffers from SPoF. To survive from this issue, the hot-standby is a implemented. In contrast, hot-standby increases network traffic and performance degradation. The rename/modify/write operation cause synchronization issues in cMDS. The clustered MDS implement fault tolerance as follows:

\begin{description}
\item{Replication: }
The replication is an alternative solution to implement fault-tolerant cMDS, but the performance of cMDS depends on the replication strategy. Each MDS in cMDS synchronizes on every update to ensure consistency \cite{dMDS}. Additionally, the synchronization can be performed periodically or upon performing update operations. The synchronization can slow down the MDS due to requirements of network accesses. Thus, the latency and network traffic is a key issue for the system. Therefore, replication is performed periodically to reduce the network accesses to enhance the performance.

\item{Hot-standby: }
The Hot-standby provides seamless service when an MDS fails \cite{dMDS}. A user unaware about the failure of an MDS and metadata is served from the hot-standby MDS. The MDS synchronizes on every update with the master and hot-standby MDS. For instance, active/active or active/passive MDS \cite{He}. Active/active standby commits the transaction only when both MDS commit. After all, this method is very costly in terms of performance, which provides strict consistency. On the contrary, active/passive MDS commits the transactions by active MDS. Passive MDS commits the transaction periodically, which introduces loose consistency.

\item{Passive mode: }
The passive mode is a standby node, used to recover an MDS, but not act as standby node. For instance, HDFS secondary NameNode stores the metadata in FSImage format and master NameNode flushes the metadata to secondary NameNode periodically. This process gets the advantages of reducing the number of network access. But the problem arises when an MDS fails. The manual recovery of MDS has to be done and takes huge hiccup times.

\item{Journaling: }
The journaling system is the most useful techniques in ensuring the data loss and data recovery. The metadata is journaled on update or periodically to ensure fault tolerance.

\end{description}

 \subsection{Hotspot Problem}
At a certain period, a file or data become popular and its access rate is exponential. The data or file that is accessed frequently by users for a day or week. The access rate increases exponentially where particular MDS are heavily loaded by metadata request and this trend remains for some period (a week, a month or a year), and then the frequency of accessing rate declines. Such kind of problem arises in Google and YouTube very frequently \cite{dMDS}. An MDS is overloaded by Millions of user requests whereas other MDS are normally loaded. The overloaded MDS struggle to serve the request on time, but it is slowed down by millions of user requests for metadata. The goal of cMDS balances the load evenly among the MDS to deal with such hotspot issue, however, only hash-based MDS can overcome the issue. The hash-based solution provides an ideal solution to this hotspot issues by placing the metadata scatteringly among the MDS. On the other hand, placement of metadata scatteringly among the MDS causes poor cache performance. This is the trade-off in designing the cMDS.

\subsection{Scalability Issue}
Undoubtedly, cMDS provides high scalability. The open challenge is to design an infinite scalable cMDS. Designing an automatic scalable cMDS is a challenge. If an MDS full, then a new MDS is elected/selected for serving metadata. Thus, incremental scalability in MDS design can solve many issues. Each MDS can hold information of many terabytes of data in cMDS. Therefore, cMDS can hold information of many petabytes of data. Therefore, augmenting new MDS can scale metadata on thousands of petabytes. Summing up, thousands of MDS can serve nearly unimaginable data size. As a consequence, clustered MDS suffers from the throughput with low-sized metadata. We are landing towards Exabytes of data. The petabytes will be obsolete in IT industries very soon. For example, Google data warehouse stores 15 Exabytes. Therefore, scalability of MDS is a major concern.

\subsection{Interoperability issue}
Interoperability is ability to exchange information in different platforms. Interoperability unlocks the restriction issues. However, the interoperability becomes a key barrier due to various reasons, namely, proprietary of the product, communication protocol, underlying architecture, etc. is the prominent issues where interoperability fails. There are no common standards for metadata operations. Therefore, an MDS is unable work for another MDS of different platform. However, the complete decoupling of metadata from data servers is a preliminary step towards the interoperability.

\subsection{Latency issue}
The latency is the most important part of MDS design, and the lowest latency is possible in stand-alone MDS. The latency is the reaction time, call for low latency, and need a fine tune for keeping it as low as possible. The cMDS incurs latency in synchronization. Precisely, the ultra-modern file system requires clustered MDS such that the file system should be scaled to Exabytes or beyond. To design such kind of MDS, the latency has been taken extra care of and it must have low latency as much as possible. RamCloud \cite{RC15,LT11} address the issues of latency problem. In another situation, the key issue of designing geo-aware cMDS is its distance. The MDS servers are located in remote areas, and its latency is very high. Neither distance can be reduced nor latency can be lowered since distance is directly proportional to latency. As distance increases, latency becomes high.

\subsection{Read/write efficiency}
\begin{description}
\item{Write Latency:}
Although, writing the files to SSD or HDD, is  inevitable, but MDS tolerates some amount of time for writing. Writing process issues distributed lock on all MDS, those are pertaining same data to synchronize the write operation. This process consumes times in seconds. Can we obviate non-volatile storage for durable metadata storage? It is literally a billion dollar question. Because keeping metadata in primary memory cause data lost. The Dr. Hadoop \cite{DR16} is the most probable answer to this question, but not durable too. On the other hand, writing process synchronizes among the MDS, as a consequence, network traffic and latency.

\item{Read latency:}
The read latency is kept as low as possible because this is the performance of the entire file system. The client issues the read operation on metadata, it should not take more network latency. The design must provide the location of metadata among MDS in the least possible time. The writing process may require more or less time, and this is tolerable, but, not reading process. Most of the queries are reading process. Once a client submits a query to cMDS, the cMDS should not take more than one network access. Unlikely, IndexFS \cite{IF14} takes more network access, causes a significant performance effect on the file system. In particular, read-efficient cMDS perform better than write-efficient cMDS, while reading metadata and vice versa as stated in Trade-off \ref{trof3}. 

\begin{trade-off}\label{trof3}
Read-efficient and write-efficient are head and tails of MDS, that never comes with a single side
\end{trade-off}.
\item{The trade-offs:}
The write-efficient MDS is based on an update, namely, modify, move, and rename. The write efficient-MDS is very useful in dynamic update performed on the file system. For example, the HDFS does not perform well in extensive write environment, due to its write-once, read-many model. The read-efficient MDS does not provide good performance in those operations except read operation. The read-efficient MDS perform very well when the update operations are rare and this model is very useful for archiving data. This is another important trade-off of designing MDS.

Additionally, the performance of the MDS decreases, when the design of MDS accommodates fault-tolerance. cMDS is highly scalable, which is not suitable for small scale metadata. The standalone MDS is very fast, not suitable for high-sized metadata. This is a trade-off of designing cMDS and it is due to communication and synchronization overhead. Some other tools can be used to store metadata, namely, Big Table \cite{BT08} and LevelDB \cite{LDB15}, but it does not work well in the large-scale metadata since these are not an intrinsic in-memory system.

Another scenario, there is a trade-off between cache-efficient MDS and load balancing. Hash-based MDS exhibits good load distributions while tree-based MDS exhibits poor load balancing. On the contrary, Hash-based MDS exhibits poor cache performance while tree-based MDS exhibits good cache performance. 

Moreover, there is a  trade-off between low latency and geo-aware MDS. Geo-aware MDS exhibits high latency while standalone MDS does not have a latency issue. Similarly, latency within a farmhouse is lower than the multiple farmhouses. However, geo-aware is most important factor in designing MDS for disaster recovery systems.

Furthermore, there is also a trade-off between strict consistency and loose consistency. Strict consistency degrades the performance while loose consistency enhances the performance of MDS. Therefore, these trade-offs are to be addressed while designing cMDS.  

\end{description}

\subsection{Adaptability issues}
The adaptability is the adjustability of any system in any situation. The MDS is decoupled from the data server. Therefore, the adjustability makes strides of MDS for any kind of database too. An MDS is adaptable only when the MDS can be used for different kind of database or file system. The adjustability requires different parameters' tuning. Consequently, the adjustability is a key issue because the unknown future use of MDS.

\subsection{Hiccup Problem}
In cMDS, metadata has to be distributed among the MDS and this distribution process needs time to become stable. In the meantime, the progress of a task may be halted due to this process. The metadata distribution must ensure minimal transfer of metadata, minimal communication, and become stable as fast as possible.
\begin{proposition}\label{prop4}
There is at least some time required to detect the faulty MDS, since there is no silver bullet
\end{proposition}

Furthermore, detection of a failure  MDS takes time. The hot-standby MDS can take over the charge of failure MDS instantly as stated in Proposition \ref{prop4}. Albeit, hot-standby exists, seconds of time required to detect failure MDS, since, there is no silver bullet. The liveliness of an MDS is detected using heartbeat. An MDS is blacklisted if the particular, MDS is unable to respond the heartbeat signal. After a few times, the MDS is marked as a dead or faulty MDS. An MDS becomes faulty or failure due to many reasons, for instance, high network traffic, network link failure, and overloaded MDS. Moreover, the problem arises on when the network partitioned among the MDS in cMDS. Even though there is CAP theorem \cite{CAP00}, still we can enhance the network partition tolerance.

\subsection{Communication Cost}
Network access introduces communication cost. Communication cost is also a grand challenge to overcome in designing cMDS. The researcher is to ensure minimization of communication cost while designing cMDS. Communication cost also depends on number of RPC and it should not be more than two RPC. An RPC can take more than a second. Thus, reducing RPC enhances the performance of cMDS drastically. The communication cost includes network traffic, data lost in the network, link failure, bandwidth and network latency \cite{dMDS}. Interestingly, one-phase commit protocol can be used to improve communication cost \cite{Congiu12,DR16}.

\subsection{Metadata Consistency}
The metadata consistency is an issue, and inconsistency is never tolerated; financial institution for instance. To maintain the consistency, network communication is performed, which takes some time. In-memory MDS does not have durability. The MDS uses a loose consistency model, namely, BASE \cite{BASE08} or Eventually Consistent \cite{EC08} model. Conventional system uses ACID model \cite{ACID81,ACID83}.

\subsection{Load balancing}
The load balancing is another issue in MDS design. The hash based has good load balancing, but, poor cache performance \cite{dMDS,MDS}. The subtree partition is not good at load balancing. The MDS should not be affected by hotspot problem. The MDS must have a capability of balancing the load dynamically.

\subsection{Migration}
Metadata migration is also a prominent issue in the MDS. The metadata is transferred from an MDS  to another MDS. The consistency and durability are ensured, before being migrated to the MDS. The migration causes readjustment in MDS, which triggers hiccup problem. The challenge is to design a metadata migration system in hash-based MDS. Interestingly, the HWM provides efficient migration technique \cite{HWM}.

\subsection{Cache Performance}
Even though the head and tail never come in a single side, there is still a good chance to enhance the cache performance in hashing. The hash-based MDS never exhibits good cache performance as well as a subtree partitioning. To make strides of cache performance, the LpH \cite{LpH94} and Consistent hashing \cite{CH97} can be used. Caching is performed using Cache memory, RAM, and HDD. Albeit, the method differs from each other. The caching technique improves performance of a system drastically. 

\subsection{Heterogeneity}
The root cause of straggler MDS is Heterogeneity. Straggler MDS unable to serve the number of MDS operations as expected. The underlying hardware configuration is always not same with all. Therefore, it is also another prominent challenge in distributed systems. cMDS cannot make an assumption of equal processing speed of all MDS. One or more MDS can also be a straggler MDS due to various reasons. For instance, low hardware configuration, high network noises, and high background noises. The straggler MDS cannot serve more amount of users while others are serving a huge amount of metadata requests \cite{dMDS}. Meantime, the users are waiting for a response in the straggler MDS. The users experience unexpected delaying of metadata services. Notwithstanding, cMDS is able to scale large amount of metadata, but is unable to serve the client requests at an expected pace. However, there is no heterogeneity issue currently in clustered MDS, but it is an issue and challenge of future in very large scale metadata size. 

\subsection{Durability}
A cMDS can crash at any time for any unforeseen reason. The cMDS stores its data in RAM, which is not permanent. Therefore, there is a need for journaling of metadata, or flushing metadata to permanent storage (disk or SSD) to guarantee the durability. Thereupon, the durability may deteriorate the performance of the MDS.

\subsection{Disaster Recovery}
The disaster recovery of any system is a challenge for past \cite{Ueno09}, present, and future \cite{dMDS}. The disaster recovery system depends on design decision of the system. Generally, cMDS run in a farmhouse where all the other servers are running. What will happen, if cMDS are damaged? What will happen, if fire caught in the farmhouse? What will happen to the data in natural calamities? Disaster recovery system requires geographical distribution of data such that it is unaffected by war, fire, earthquake, flood and any other natural calamities. Therefore, performance of cMDS may degrade due to disaster recovery management system, however, the data are unaffected by any disaster. Notably, disaster recovery system is costly. The challenge is to design disaster recoverable cMDS without affecting the cMDS performance.

\section{Discussion}
\label{dis}
Distributed systems interconnect the MDS through low-speed networks and uses shared-nothing architecture. Unlike distributed system, the parallel systems are interconnected with high-speed networks and uses shared everything architecture. LustreFS \cite{LustreFS15}, for example. Even though their architecture is different, but, both systems deal with large-scale metadata without any failure. The most of the cMDS are dealing with fault-tolerance system, namely, Dr. Hadoop \cite{DR16}, DROP \cite{DP13}. The fault-tolerance is attained by replication, journaling, erasure code, RAID, and de-duplication. Also, most systems depend on sMDS, which is the most prone to failure. We have found that many parameters of cMDS are overlooked in designing MDS. For example, what are the constructive measures, if the MDS crashes? A cMDS implements hot-standby or any failover mechanism. Consequently, the communication cost becomes a barrier for the cMDS. For example, Dr. Hadoop \cite{DR16} uses one phase commit protocol \cite{Congiu12} to reduce the communication cost. The cMDS adapts counter measures for faulty MDS. The fault-tolerance is the most important factor to take care of, since most of the metadata are kept in RAM. Therefore, we must recall a key quote in designing cMDS as \textit{``Everything is lost by losing metadata''}. But, there is a dark side of the cMDS too, for instance, fails in throughput, and suffers from synchronization. Notably, the cMDS is able to serve the metadata on a scale of infinite, claimed by Dr. Hadoop \cite{DR16} for instance. 

The geo-aware MDS is capable of handling disaster management of metadata. The metadata is replicated in multiple geo-location \cite{Liu} and it is fetched back if a disaster happens. However, the geo-aware MDS is suitable for the high-risk data management system, e.g., data of financial institution. The CalvinFS \cite{CF15} has big advantages over the traditional MDS. The CalvinFS provides WAN replicated MDS and it replicates the metadata in geographical location. This is replicated based cMDS and therefore, there is a latency issue. In this design, the disaster recovery is features and a possible event, but, network partition cannot harm the file system. If the disaster is not the concern, then this design gives high latency. Similarly, the MRFS \cite{Yu14} is also designed to overcome the problem of disaster.

\section{Conclusion}
\label{con}
In this article, we have discussed important trade-off between a) hash-based and tree-based MDS,  and b) read-efficient and write-efficient MDS. In addition, clustered MDS and standalone MDS are exploited. The article provides insight on available based on the MDS design methodology. The issues and challenges of MDS are also highlighted. 

As we have discussed, the metadata size is growing, and thus, the standalone MDS is nearly obsolete. Therefore, the clustered MDS emerges. The clustered MDS is categorized into two key categories dMDS, and pMDS. The various available methods encourage to design a good MDS and creates new possibilities in designing cMDS. On the contrary, the cMDS designs can have one or more issue(s), namely, durability, consistency, latency, network traffic, heterogeneity, scalability, fault-tolerance, load-balancing, cache performance, and disaster recovery. The standalone MDS and clustered MDS (cMDS) is applicable as per the demand of the situation. The metadata resides in RAM, and therefore, an MDS requires in-memory data structures. In this survey, we have found that there are dozens of parameters to be considered till now, which has been continuously overlooked. Moreover, we have shown all issues of the cMDS design which are very helpful to the research community in designing of ultra-large cMDS design. The careful designing of cMDS helps in delivering the promising storage system. Finally, the authors firmly believe that this study will serve as a major milestone in the designing of upcoming cMDS design.

\bibliographystyle{ACM-Reference-Format}
\bibliography{sample-base}










\end{document}